\documentclass[aps,pre,twocolumn,amsmath,amssymb,amsfonts]{revtex4}
\usepackage{epsfig}
\usepackage{graphicx}
\usepackage{dcolumn}
\usepackage{bm}
\begin{document}

\title{Refined central limit theorem and infinite density tail of the Lorentz gas from L\'{e}vy walk}

\author{Itzhak Fouxon$^{1,2}$}\email{itzhak8@gmail.com}
\author{Peter Ditlevsen$^{2}$}\email{pditlev@nbi.ku.dk}

\affiliation{$^1$ Centre for Ice and Climate, Niels Bohr Institute, University of Copenhagen, Copenhagen, Denmark}
\affiliation{$^2$ Department of Computational Science and Engineering, Yonsei University, Seoul 120-749, South Korea}

\begin{abstract}

We consider point particle that collides with a periodic array of hard-core elastic scatterers where the length of the free flights is unbounded (the infinite-horizon Lorentz gas, LG). The Bleher central limit theorem (CLT) states that the distribution of the particle displacement divided by $\sqrt{t\ln t}$ is Gaussian in the limit of infinite time $t$. However it was stressed recently that the slow convergence makes this result unobservable. Using a L\'{e}vy walk model (LW) of the LG, it was proposed that the use of a rescaled Lambert function instead of $\sqrt{t\ln t}$ provides a fast convergent, observable CLT, which was confirmed by the LG simulations. We demonstrate here that this result can simplified to a mixed CLT where the scaling factor combines normal and anomalous diffusions. For narrow infinite corridors (almost finite-horizon case) the particle for long time obeys the usual normal diffusion, which explains the previous numerical observations. In the opposite limit of small scatterers the Bleher CLT gives a good guiding observable, and can be used with about fifteen per cent accuracy. In the intermediate cases the mixed CLT applies. The obtained Gaussian peak determines moments of order smaller than two. In contrast, the CLT cannot fully describe the coordinate dispersion, of which it only gives half in the long time limit, and also moments of order higher than two. These moments and the missing half of the dispersion are described by the distribution's tail (the infinite density) which we derive here. The tail is supported along the infinite corridors and formed by anomalously long flights whose duration is comparable with the whole time of observation. We reinforce the relevance of one flight, however large time is, by demonstrating that the tail and the moments of order higher than two depend on whether the particle moves ballistically between the steps or instantaneously jumps at the end of each step as in the usual random walk. The moments' calculation from the tail is confirmed by direct calculation of the fourth moment from the statistics of the backward recurrence time defined as time that elapsed since the last collision. This completes the solution of the LW model allowing full comparison with the LG.

\end{abstract}

 \maketitle

\section{Introduction}

The Lorentz gas (LG) is one of the most studied systems in the theory of chaos and non-equilibrium statistical physics, see e. g. \cite{lor,hard,Dettmann,chre,gasp} and references therein. In a simplest formulation the system consists of a pointlike particle that moves at unit speed and undergoes instantaneous elastic collisions with non-overlapping hard-core circular scatterers arranged in a $2d$ periodic lattice.  The LG describes the simplest non-trivial mechanical system of two hard disks colliding in a finite volume with periodic or reflecting boundary conditions \cite{Dettmann}. It also provides a simple model for the interaction of light and heavy particles usable in the study of conductivity \cite{cond}. Recently the model became of use in artificial graphene albeit with a soft potential \cite{kals}.

From the theoretical viewpoint the LG is one of the unique systems where the macroscopic laws can be derived from the microscopic ones. A main object of interest is the statistics of the particle's displacement from the initial position $\bm r(t)=\int_0^t \bm v(t')dt'$ where $\bm v(t)$ is the particle velocity. We would like to know whether the statistics becomes Gaussian at large times, corresponding to asymptotic diffusion, and, if it does not, which statistics comes instead. Here the statistics is defined by assuming that initial conditions are distributed according to the equilibrium microcanonical ensemble modulo lattice translations (for reduction to variables on the boundary, see e. g. \cite{fma84}). Random distribution of scatterers' positions can also be used for introducing the statistics \cite{dett14}.

In the case of the finite-horizon LG, where the free path is bounded, convergence of the probability density function (PDF) of $\bm r(t)/\sqrt{t}$ at $t\to\infty$ to a Gaussian distribution (CLT) is proved in \cite{sb}. Thus the PDF of $\bm r(t)$ obeys at large times the diffusion equation (with possibly matrix diffusion coefficient). This result can be understood by observing that the lower bound on the number of collisions $N(t)$ in time interval $t$ grows with $t$ linearly. Thus the decay of the memory of the direction of the initial velocity, that must hold after several collisions with the convex scatterers, implies the memory decay in time. The velocity autocorrelation function $C(t)$ obeys stretched exponential decay \cite{sb,bdec,decay,bleher}. This implies that $\int_0^t \bm v(t')dt'$ at large $t$ is roughly a sum of large number of independent random variables producing the CLT. An example of the finite-horizon LG is provided by a triangular lattice \cite{dett14}. In contrast a square lattice with non-overlapping scatterers produces unbounded free paths. For instance a lattice with scatterers' radius between
$1/\sqrt{8}$ and one half (in units of lattice constant) forms infinite corridors parallel to the coordinate axes in which the particle could move without collisions \cite{det,us}. This case, which is considered in this work and called the infinite-horizon LG, is still not fully understood. The most prominent open problems are inobservability (at least, as of today) of the leading order behavior at large times and the issuing need for the description of the finite-time behavior, and the tail of the distribution that determines moments of order higher than two. Description of the problems and the progress brought by the present work in their solution demand quite a lot of details and is done in a separate Section below.

\section{Open problems of the infinite horizon LG and our main results} \label{open}

The infinite-horizon LG exhibits superdiffusive behavior: the finite-time diffusion coefficient $D(t)=\langle r^2(t)\rangle/(4t)$ diverges at large times logarithmically (here and below the angular brackets stand for average) \cite{fma84,zach,bleher,dahl}. This divergence is seen simpler by considering the evolution in terms of the number of collisions rather than the continuous time \cite{bleher}. The transformation is done using the law of the large numbers according to which with probability one $\lim_{t\to\infty} t/N(t)$ equals the mean free time $\langle \tau\rangle$. We find the asymptotic equality $D(t)\sim D_{N(t)}/\langle \tau\rangle$ where the discrete diffusion coefficient $D_N$ is defined by the moment $t_N$ of the $N-$th collision as $D_N\equiv \langle r^2(t_N)\rangle /(4N)$.

Finiteness of the first moment of the free time between the collisions $\tau$, used above, holds since the PDF $\psi(\tau)$ of $\tau$ has a cubic tail \cite{bleher,bouch,zach,us}.
It is this fat tail, formed by the long flights, that causes the anomalous diffusion properties of the infinite-horizon LG. Indeed, writing $\bm r(t_N)=\sum_{i=1}^N \bm \Delta_i$ where $\bm \Delta_i$ is the displacement vector between collisions $i-1$ and $i$ (with $\bm \Delta_1\equiv \bm r(t_1)$) we find \cite{bleher},
\begin{eqnarray}&&\!\!\!\!\!\!\!\!\!\!\!\!\!\!
\lim_{N\to\infty}D_N=\frac{\langle \Delta_1^2\rangle}{4}+\frac{1}{2}\sum_{k=1}^{\infty}\langle \bm \Delta_1\cdot\bm \Delta_{k+1}\rangle.
\end{eqnarray}
The correlation functions in the sum are finite and decay quickly with $k$ (the convexity of the scatterers causes fast decay that dominates the possibly large magnitude of the steps). The divergence of $D_N$ and $D(t)$ is caused by the divergence of $\langle \Delta_1^2\rangle=\langle \tau^2\rangle$, observed in \cite{fok}, that is, by single long flights rather than long correlations of different steps.
Despite that the flights are rare they increase the displacement so much that their contribution in $\langle r^2\rangle$ cannot be neglected.

The divergence of $\langle \Delta_1^2\rangle$ is logarithmic so $D(t)\propto \ln t$ must hold at large times. Since by the Einstein-Green-Kubo formula the diffusion coefficient is proportional to the time integral of $C(t)$ then $C(t)\propto t^{-1}$ holds \cite{fma84,bdec}. These long correlations are the reason why for the infinite-horizon LG $\int_0^t \bm v(t')dt'$ cannot be considered as a sum of large number of identically distributed random variables.

The $\langle r^2(t)\rangle \propto t\ln t$ behavior implies that the PDF of $\bm r(t)/(\sqrt{t\ln t})$ might have a finite $t\to\infty$ limit. This is indeed the case as was proved by Bleher \cite{bleher} who demonstrated that the limiting distribution is Gaussian. Unfortunately this modified CLT is less powerful than its finite-horizon counterpart. There seem to be two major points demanding the refinement of the result.

The first point concerns the calculation of the asymptotic behavior of the moments of the displacement at large times. Convergence in distribution does not imply the convergence of the moments. This difference is insignificant in the finite horizon case where the moments can be calculated from the Gaussian PDF, however in the infinite horizon case this is not necessarily the case. The calculation of the dispersion for the limiting distribution gives only half the true value \cite{det,chern}. The dispersion is contributed equally by the most probable diffusive motions that determine the Gaussian center of the distribution of $\bm r(t)$, described by the CLT, and the long flights that determine the distribution's tail \cite{det}. Therefore the calculation of the dispersion from the PDF demands the knowledge of the tail.
Moreover the dispersion is a borderline moment \cite{det}. The moments with order smaller than two can be found using the Gaussian center of the distribution however the higher-order moments are fully determined by the tail. For these moments the Gaussian formula is wrong by a large parameter. This phenomenon, called intermittency, holds because the most probable events, that determine the Gaussian central part of the PDF, contribute these moments negligibly. The moments are determined by rare long flights in infinite corridors (if the flight is long then it occurs in the infinite corridor \cite{bleher}) that form the PDF's tail. The necessity of knowing the tail is evident from the structure of trajectories as seen in the simulations, see e. g. \cite{us}. The growth of the displacement typically consists of long periods of nearly normal diffusion, intermitted by rare single event flights that take the particle so far that they provide significant contribution in the statistics of the displacement. The distribution's tail is not available today.

The other problem of the CLT can be seen by considering the transition from the finite to the infinite horizon case, that occurs when the dependence on the lattice geometry is studied. Thus for a square lattice with unit lattice constant and circular scatterers of radius $R$ the transition occurs at $R=1/2$. It can be called "opening transition" since when $R$ crosses one half from above, infinite horizons open for the first time. For infinitesimally small positive $1/2-R$ the LG has a nearly finite horizon. A long flight along one of the infinite corridors is such a rare event that on finite time-scales the LG is indistinguishable from the finite horizon. The ordinary CLT for the PDF of $\bm r(t)/\sqrt{t}$ must give a good approximation.
At the same time in the infinite time limit, the Bleher CLT must hold. Here we demonstrate that roughly a simple interpolation between the results holds, that can be called a refined CLT. That states in the next to leading order at large times the PDF of $\bm r(t)/(\sqrt{t(\ln t+2\zeta(R)})$ is approximately Gaussian. The leading order approximation at large times is the Bleher CLT, however $\zeta(R)$ diverges at $R=1/2$ so that CLT is unobservable at $1/2-R\ll 1$ (here the factor of two in front of $\zeta$ is introduced for correspondence with the formulas below). The problem exists also at $1/2-R\sim 1$ where the Bleher CLT demands $\ln\ln t\ll \ln t$, as seen from \cite{bleher} and stressed in \cite{us}. This condition never holds at realistic times unless one is ready to put up with about twenty per cent accuracy. The refined CLT, introduced in \cite{us}, encompasses all these cases. Here we provide a reduction of the result of \cite{us} and also a separate consideration of the instructive discrete case. The refined CLT is indispensable in observations where the Gaussianity of $\bm r(t)/(\sqrt{t\ln t})$ might be unobservable \cite{us}. These considerations are in agreement with the previously observed properties of the LG's dispersion, considered in more detail below, where the normal diffusion of the infinite horizon LG was observed to dominate the superdiffusive behavior sometimes \cite{cr1}.

The refined CLT, proposed using the LW model of the LG in \cite{us}, has Lambert function instead of $\ln t+2\zeta$ above. This CLT was demonstrated numerically to hold for the Lorentz gas at $R=0.4$ starting from times obeying $N(t)\sim 10^4$ when the Bleher CLT is invalid. The reduction proposed here provides a simpler form of this result and demonstrates that the main reason for the unobservability of the Bleher CLT in \cite{us} was a large value of $\zeta\simeq 9$ holding for the considered $R=0.4$. The $R=0.4$ is the case of a mixed CLT where Gaussianity holds due to the largeness of the sum $\ln t+2\zeta$ (and the next order corrections) and not of one of the summands separately. Thus both normal and anomalous diffusive events form the Gaussian peak of the PDF. We demonstrate that due to the quadratic divergence of $\zeta(R)$ at $R=1/2$, for smaller $1/2-R$ the refined CLT provides for a long period of normal diffusion before the change to the Bleher CLT occurs at the largest times. In contrast, decrease of $R$ leads to a unlimited growth of the number of infinite corridors and decrease of $\zeta(R)$ to values less or of order one, implying observability of the Bleher CLT, corrected by $\ln\ln t$ term. Here the decrease of $R$ produces two more infinite corridors at $R=1/\sqrt{8}$ and further decrease eventually leads to divergence of the number of infinite corridors at $R\to 0$. For the moment, the transfer of the refined CLT, proved rigorously in the frame of the LW model, to the LG can be considered as proved for $R=0.4$ making it highly plausible that it holds at other $R<1/2$ also. Generalization to other lattice geometries, such as triangular \cite{bleher}, is possible however is beyond the scope of this paper.

In this work, besides the described progress on the structure of the refined CLT, we derive the tail of the PDF of the LW. Testing this form for the LG, where currently the tail was not reached in the simulations (see e. g. \cite{us}), would be of high interest.

We use the LW model that was introduced recently in \cite{us}. The model utterly neglects the correlations of $\bm \Delta_i$. Thus the motion is a random walk with independent steps. This assumption is not so unreasonable. It was observed in \cite{bleher} that the correlation coefficient $\langle \bm \Delta_i\cdot \bm \Delta_k\rangle/\sqrt{\langle \bm \Delta_i^2\rangle\langle \bm \Delta_k^2\rangle}$ is zero for $i\neq k$ since the numerator is finite and the denominator is infinite. Thus effectively $\Delta_i$ with different indices are independent, though making rigorous sense of this observation demands quite difficult considerations \cite{bleher}. Further, the model introduces the natural decomposition of the displacement $\Delta_i$ into the product of the constant random velocity of the $i-$th step of the walk, $\bm v_i$, and the $i-$th step's duration, $\tau_i$. This decomposition is needed for passing from discrete steps to the continuous time variable. A "molecular chaos" type assumption is made that the constant velocity $\bm v_i$, which is determined by the previous collisions of the particle, is independent of $\tau_i$, which is determined by the next collision. No further approximations are made for the distribution of $\tau$ for which the "true" LG's PDF $\psi(\tau)$ is used. Thus $\psi(\tau)\propto \tau^{-3}$ holds at large $\tau$ where the correction is of order $\tau^{-7/2}$, see \cite{bleher} (the usage of the correction is significant in some questions, slightly differing our study from \cite{us}). We remark that the random walk model might also be of use in the finite-horizon case, not considered here, where $\psi(\tau)$ has a compact support.

The LW model assumes that the PDF of the velocity $F(\bm v)$ obeys $F(\bm v)=F(-\bm v)$ since the PDF of $\bm \Delta=\bm v\tau$ of the LG is symmetric \cite{bleher}. The form of the PDF of the velocity $F(\bm v)$ is fixed uniquely from the geometry of the tail. For large $\Delta$ the PDF of $\bm \Delta=\bm v \tau$ is localized near the directions of the corridors \cite{bleher} and thus this shape must be repeated by $F(\bm v)$. In the case of only two corridors perpendicular to each other, see above, this dictates the distribution of the velocity where the particle at each step goes with equal probability and velocity to the left, to the right, up or down,
\begin{eqnarray}&&\!\!\!\!\!\!\!\!\!\!\!\!\!\!
F(\bm v)=\frac{V\left(\delta(v_y)\delta(v_x^2-V^2)+\delta(v_x)\delta(v_y^2-V^2)\right)}{2},\label{vela}
\end{eqnarray}
where $V$ is the velocity magnitude, set below to one. It is readily seen that the usage of this $F(\bm v)$ must produce realistic PDF $P(\bm r, t)$ of $\bm r(t)$ at large times. Indeed, the structure of the tail of the PDF $P(\bm r, t)$ repeats that of the tail of the PDF of the single step $\bm \Delta$. Thus $P(\bm r, t)$  takes a cross-like shape by growing fat tails along the directions of the corridors parallel to $x$ and $y$ axes \cite{us}. At the same time, the usage of this $F(\bm v)$, despite that it is unrealistic for moderate $\Delta$, would still produce a realistic center of the PDF. This is because the center is formed by many steps and is insensitive to geometry of one step similarly to the traditional random walk (see however \cite{2016,2017} where sensitivity of the PDF's center to the structure of one step of the random walk was demonstrated for the LW with slower than cubic decay of $\psi(\tau)$). These considerations are confirmed by the LG simulations that demonstrate that $P(\bm r, t)$, found with this $F(\bm v)$, agrees perfectly with the data without any fitting \cite{us}. This finding did not include the farthest tail of the PDF that describes the PDF's decay toward the maximal possible displacement $Vt$, due to the numerical constraints and because the theory of \cite{us} concentrated on the PDF's center. The tail is necessary for the full description as detailed above. Our derivation of the tail here is either done for a general space dimension $d$ or can be generalized trivially. Since the cubic tail of $\psi(\tau)$ holds for the LG with $d<6$, see \cite{det,us}, then the results apply also in the three-dimensional case.
We also use a general fast decaying $F(\bm v)$ having in mind uses with other geometries and/or coarse-grained descriptions defined below.

We observe that the possibility of describing the LG by a continuous time random walk (CTRW) is in fact true rigorously, cf. \cite{bg}. The strong ergodic properties imply that correlations of $\bm \Delta_i$ and $\bm \Delta_k$ rapidly decay to zero at $|i-k|\to\infty$ and the corresponding $\bm \Delta_i$ and $\bm \Delta_k$ can be considered as independent \cite{bleher}. Thus the coarse-grained displacements $\bm \Delta'_i$ and $\bm \Delta'_{i+1}$ over $n$ free flights, $\bm \Delta'_i=\sum_{k=in+1}^{(i+1)n}\bm \Delta_k$ can be considered as independent for a certain finite $n$ (a simultaneous limit of infinite $n$ and number of collisions $N$ with $n\ll N$ is needed in a formal proof). This gives $\bm r(t_N)$ for large $N$ as the sum of independent random vectors $\bm \Delta'_i$ producing a CTRW. The total duration $\tau'_i$ of one coarse-grained displacement is a sum of $n$ free times and it has the cubic tail deriving from $\psi(\tau)$. The precise form of the PDF of $\tau'_i$ is irrelevant since most of the properties (asymptotic scaling of the moments in time) depend on the existence of the cubic tail only. The main difference from the LW is in the interdependence of the effective velocity $\bm v'_i=\bm r'_i/\tau'_i$ and $\tau'_i$. For very large $\tau'_i$ one flight determines the whole sum $\bm  r'_i$ and the magnitude of velocity $\bm v'_i$ is close to unit velocity of the original LG. In contrast for typical $\tau'_i$ the magnitude of $\bm v'_i$ can differ much from one. The interdependence complicates the derivations and could produce a whole range of diverse scaling laws \cite{ctr}. However the simulations of \cite{us} indicate that in our case the LW described in the previous paragraph applies and we will assume its validity as an empirical fact.
Despite the incompleteness in the LW's introduction, it seems highly plausible that the LW reproduces all the qualitative properties of the LG, both known and unknown, see examples below.

The LW, assuming its validity in the LG's description, gives a simple way to understanding the LG and developing further knowledge of it. The LW is solved in quadratures: the Montroll-Weiss equation \cite{UchZol,Zaburdaev,2017} gives the Fourier-Laplace transform of the displacement's PDF via $F(\bm v)$ and $\psi(\tau)$. Thus the asymptotic evaluation of the inverse transform in the large-time limit can be used for deriving
the asymptotic scaling laws, studying the universality and considering other questions. The diffusion coefficient and the CLT, with the width of the Gaussian peak given by half the dispersion, are easily derivable, see \cite{us} and below. The results previously unknown for the LG can be derived for the LW and then tested on the LG.

Probably the most robust and at the same time rather simple prediction of the above LW is,
\begin{eqnarray}&&\!\!\!\!\!\!\!\!\!\!\!
\langle r^2(t)\rangle= \frac{2\left\langle v^2\right\rangle A_1t}{\langle \tau\rangle} \left(\ln t+\zeta-2+C+O\left(t^{-1/2}\right)\right), \label{growth}
\end{eqnarray}
where $A_1$ and $\zeta$ are constants characterizing the geometrical properties of the lattice of the scatterers and $C=0.577215$ is the Euler's constant. Below we call $\zeta$ the crowding parameter since it characterizes how dense the scatterers are around the particle, becoming infinite when the infinite corridors become very narrow, and tending to zero for small scatterers sparsely distributed in space. This parameter corrects the 
logarithmic growth of the diffusion coefficient, which is proportional to the term in brackets, by a constant. The derivation of Eq.~(\ref{growth}) was done in the SM of \cite{us}. Here we make the significant observation that the next order correction is of order $t^{-1/2}$ and is not logarithmic, see Sec. \ref{if}. Since the condition of the negligibility of the correction, $t^{-1/2}\ll 1$ is mild, then the finite-time diffusion coefficient must be well represented by a linear combination of the logarithmic (superdiffusive) and constant (normal diffusion) terms.  In contrast, the condition of observability of the purely logarithmic growth, which is that the logarithm dominates the constant term, is rather difficult to realize unless $\zeta\lesssim 1$ (that holds for $R$ significantly below $1/2$ see below). This fully fits the results of the LG simulations of \cite{cr1}. The authors found that the logarithmic asymptotic growth cannot be attained in their simulations however inclusion of the normal diffusion correction provides accurate description of the data. Moreover it was found that the normal diffusion term may dominate the logarithmic growth. The reason was traced to the observation that if Eq.~(\ref{growth}) is valid, then the constant in the RHS must diverge in the limit of narrow corridors. Thus for square lattice, $\zeta$, considered as the function of the radius $R$ of the scatterers, must diverge at $R\to 1/2$. We demonstrate here, using the results of \cite{us}, that the divergence is quadratic, as we already remarked previously. When $1/2-R$ is finite the constant $\zeta$ still can be large. For instance $\zeta \simeq 9$,  obtained in \cite{us} for $R=0.4$, is rather large on the logarithmic scale. This results in the existence of an intermediate normal diffusion regime where $t^{1/2}\gg 1$, necessary for the validity of Eq.~(\ref{growth}), holds but $\ln t$ is still smaller than $\zeta-2+C$. Thus the LW's prediction for the behavior of the LG diffusion coefficient given by Eq.~(\ref{growth}) is in agreement with the LG data. Whether the agreement is also quantitative demands comparison with the simulations that are outside our scope here.


It is of high interest if the refined CLT of \cite{us} can be derived directly from the LG mechanics. We give in Sec. \ref{fast} a strong indication that indeed the refined CLT can be proved for the LG. We observe that the proof of the CLT for the LG \cite{bleher} relies on the observation that the distribution of $\sum_{i=1}^N \bm \xi_i/\sqrt{N\ln N}$ at $N\to\infty$ is Gaussian, see also \cite{pr}. Here $\bm \xi_i$ are independent identically distributed random variables whose distribution coincides with that of $\bm \Delta$, cf. Khintchine-Feller-L\'{e}vy theorem  \cite{gned,bg}. This observation does not rule out the possibility that there is a factor $A_N$ so that the distribution of $\sum_{i=1}^N \bm \Delta_i/A_N$ converges to the Gaussian distribution faster in $N$. Despite that at largest $N$ this factor must be proportional to $\sqrt{N\ln N}$ it can be quite different at a finite $N$. Indeed, we demonstrate in Sec. \ref{fast} that $A_N$ proportional to Lambert function appears in this setting naturally. The use of this result in the considerations of \cite{bleher} would produce the refined CLT for the LG though the detailed proof is outside the scope of the present paper.

We observe that the considered $\nu=2$ case of the LW with the tail $\psi(\tau)\propto \tau^{-1-\nu}$ is at the border between the normal diffusion and the superdiffusion.  If $\nu>2$ then $\langle \Delta^2\rangle<\infty$, the usual CLT holds and $\langle r^2(t)\rangle\propto t$ is found from the Gaussian center of the PDF. In contrast, if $1<\nu<2$ then $\langle \Delta^2\rangle=\infty$ and the superdiffusive growth of $\langle r^2(t)\rangle\propto t^{2/\nu}$ is determined by the tail of the PDF. The PDF in the superdiffusive case is non-universal in dimension higher than one \cite{2016,2017}. The shape of the PDF's center is determined by a function characterizing the microscopic evolution and there is no symmetry restoration at large times and scales. This poses the question of how the usage of $F(\bm v)$ other than that in \cite{us} would change the shape of the PDF's center, which is isotropic in the considered case. This question is left for future work.

Our LW calculation of the farthest tail of the PDF provides the PDF of $\bm r(t)$ near the cutoff at $r=t$ (we set $V=1$). This cutoff is the maximal displacement which is reached by the particle that moves with the unit velocity without collisions during the whole time of the observation $t$. The tail is given by a superposition of two power laws. Its form implies that the integrals for the moments of order less than two diverge at small distances, whereas those for the moments of the order higher than two diverge at large distances. The name "infinite density" for the tail refers to the divergence of the normalization integral (moment of order zero) at small distances. This density was introduced previously for the L\'{e}vy walks with slower than cubic decays of the PDF of $\tau$, see e. g. \cite{BarkaiPR,BarkaiPhysRev,BarkaiPhysLett}.

The behavior of the integral for the moments weighted by the infinite density implies that the moments of order smaller than two are provided by the central part of the PDF described by the refined CLT. In contrast, the integrals for the moments of order higher than two are determined by the events with $r(t)\simeq t$ that is events where the particle made a flight(s) that lasts for the time comparable with the whole time interval $t$. This is true however large $t$ is, despite the decrease of the probability of these flights with $t$. Finally the second moment (dispersion) is contributed both by the diffusive trajectories forming the central part of the PDF and by the long flight events. For this moment there is no typical event that forms it. The refined CLT and the infinite density together allow to find all the moments and provide quite a complete description of the displacement statistics.

Finally we illustrate the statistical significance of one step of the walk by comparing the results of the LWs where the particle moves ballistically during the walk's steps (sometimes called velocity model) and jumps at the end of each step similarly to the usual diffusion on the lattice (jump model). For the usual random walk the difference between the models is negligible (details of one step are irrelevant in the long-time, large-scale limit) but for the L\'{e}vy walk the situation can be different, see e. g. \cite{BarkaiPR}. We demonstrate that the dispersions of the two models differ by one half in the constant term in the brackets of Eq.~(\ref{growth}). This difference is negligible at $\ln t+\zeta\gg 1$ and thus can be disregarded (the condition $t^{1/2}\gg 1$ of validity of Eq.~(\ref{growth}) guarantees that the difference is negligible compared with $\ln t$). However the infinite density tails are "footprints" of $F(\bm v)$ and differ for the models. For the velocity model the PDF does not vanish at $r=t$, however for the jump model it vanishes since the particle does not move until the step's end. This results in the significant difference of the factor $\beta/2$ for the moments $\langle r^{\beta}(t)\rangle$ with $\beta>2$. One step matters for these moments.

The text below is organized in this way. In the next Section we introduce the definition of the LW and the Montroll-Weiss equation. Section \ref{if} considers the finite-time diffusion coefficient. Section \ref{fast} provides the refined CLT for the discrete sum of variables demonstrating how the Lambert function appears and reduces to the sum of normal and anomalous diffusion scalings. The next Section provides the refined CLT for the continuous time LW by a reduction of the result of \cite{us}. Section \ref{sum} derives the dispersion of the jump model and compares it with the velocity model. The next Section derives the fourth moment of the displacement using the backward recurrence time (time since the last collision), an object of its own interest for the LG.  Section \ref{ifns} derives the infinite density tail preparing the ground for the calculation of all the moments in Section \ref{ifns}. We compare the infinite density tails and high-order moments of the velocity and jump models in Section \ref{tailk}. The Conclusions resume our findings only shortly since the resume is provided in this Section. For transparency in the beginning of each Section we describe the results obtained in that Section.

\section{L\'{e}vy walk model of the Lorentz gas}

In this Section we introduce the LW model of the LG. The motion consists of a sequence of independent flights of random duration $\tau_i$ where $i$ is the flight's index. During each flight the particle's velocity $\bm v_i$ is constant. Upon the end of each flight both the velocity and the duration of the next flight are randomly refreshed. These are drawn independently with the PDFs $F(\bm v)$ and $\psi(\tau)$.

The function $\psi(\tau)$ is taken as the distribution function of the free time of the infinite-horizon Lorentz gas.
This function cannot be described completely however the asymptotic laws of propagation at large times depend on several robust quantities. The results of \cite{bleher} imply that the tail of $\psi(\tau)$ obeys,
\begin{eqnarray}&&\!\!\!\!\!\!\!\!\!
\psi(\tau)= \frac{2A_1}{\tau^3}+O\left(\tau^{-7/2}\right),
\end{eqnarray}
with a certain positive constant $A_1$. In the case of two-dimensional square lattice of circular scatters of radius $R$ this formula holds at $R<1/2$ where distances are measured in units of the lattice constants. The cubic tail must disappear when $R$ approaches $1/2$ from below which is described by the observation of \cite{us} that,
\begin{eqnarray}&&\!\!\!\!\!\!\!\!\!
A_1=\frac{(1-2R)^2}{\pi R}, \label{as}
\end{eqnarray}
holds at $1/\sqrt{8}<R<1/2$. This formula guarantees that the resulting LW's prediction for the dispersion, given by Eq.~(\ref{growth}),
\begin{eqnarray}&&\!\!\!\!\!\!\!\!\!\!\!
\lim_{t\to\infty}\frac{\langle r^2(t)\rangle}{4t\ln t}= \frac{\left\langle v^2\right\rangle A_1}{2\langle \tau\rangle}=\frac{(1-2R)^2}{2\pi R \langle \tau\rangle},
\end{eqnarray}
reproduces the LG formula of \cite{bleher}, where we took $\langle v^2\rangle=1$ since in the LG $v^2=1$ is the conserved energy. Thus the cubic tail must disappear at the opening transition at $R=1/2$ does so   
quadratically in $R-1/2$, cf. \cite{bouch}. Dependence of $A_1$ on $R$ changes at $R<1/\sqrt{8}$ where besides the infinite corridors in $x$ and $y$ directions there are other infinite corridors. We have at small $R$ that $A_1\propto R^{-2}$, see \cite{bleher}.

The above implies that the Laplace transform,
\begin{eqnarray}&& \!\!\!\!\!\!\!\!\!\!\!\!\!\!
\psi\left(u\right)\equiv \int_0^{\infty} \exp\left(-u\tau\right)\psi(\tau)d\tau,
\end{eqnarray}
obeys the asymptotic expansion,
\begin{eqnarray}&&\!\!\!\!\!\!\!\!\!
\psi(u)=1-\langle \tau\rangle u-A_1u^2\ln u+A_2u^2+O(u^{5/2}), \label{sml}
\end{eqnarray}
where $A_2$ is a constant. Proper non-dimensionalization of the argument of the logarithm can be done using $\ln \left(\langle \tau\rangle u\right)$ instead of $\ln u$ and similarly for the formulas below. We will keep the tradition of having formulas that contain logarithms of dimensional quantities \cite{bleher} where in the final answer dimensions can be restored. The $u^2\ln  u$ term in Eq.~(\ref{sml}) can be seen by considering the small $u$ behavior of the third derivative of $\psi(u)$,
\begin{eqnarray}&&\!\!\!\!\!\!\!
\frac{d^3\psi(u)}{du^3}=-\int_0^{\infty} \psi(\tau)\tau^3\exp[-u\tau] d\tau=-\int_0^{\infty} \frac{d\tau}{u}   \\&&\!\!\!\!\!\!\!\exp[-u \tau] \frac{d}{d\tau}[\psi(\tau)\tau^3]\!\sim\! -\frac{1}{u}\int_0^{\infty} d\tau \frac{d}{d\tau}[\psi(\tau)\tau^3]\!=\!-\frac{2A_1}{u},\nonumber
\end{eqnarray}
where the dots stand for solution of $\psi^{(3)}=0$ that is parabolic function. We keep the next order, quadratic in $u$ term in Eq.~(\ref{sml}) because it remains finite when $R$ approaches $1/2$ from below, in contrast with "leading" order $u^2\ln u$ term that disappears at $R=1/2$.
This term describes the contribution of the normal diffusion events as can be seen by considering the representation of the crowding parameter, $\zeta\equiv A_2/A_1$,
implied by the formulas of \cite{us} (who used a different notation),
\begin{eqnarray}&&\!\!\!\!\!\!\!
\zeta=\frac{3}{2}-C+\lim_{T\to\infty}\left(\frac{1}{2A_1}\int_0^T \psi(\tau)\tau^2 d\tau-\ln T\right),
\end{eqnarray}
where $C=0.577215$ is Euler's constant. We introduce a (non-unique) constant $T_0$ so that $\psi(\tau)\approx 2A_1/\tau^3$ for $\tau>T_0$ and the events with $\tau<T_0$ can be considered as "normal diffusion events". Indeed, it was observed in \cite{bleher} that if the particle moves for long time without collisions then it moves in one of the infinite corridors, so that the separation in normal collisions with nearby scatterers and motions in the infinite corridors is meaningful. Then we can write ($A_2=\zeta A_1$),
\begin{eqnarray}&&\!\!\!\!\!\!\!
A_2=A_1\left(\frac{3}{2}-C\right)+\frac{1}{2}\int_0^{T_0} \psi(\tau)\tau^2 d\tau
-A_1\ln T_0,
\end{eqnarray}
where $T_0$ can be considered as independent of $R$ at $1/\sqrt{8}<R<1/2$. We find that $A_2\approx \int_0^{T_0} \psi(\tau)\tau^2 d\tau/2$ which is the dispersion of the flight times due to normal diffusion events. Thus at $1/2-R\ll 1$ we can write,
\begin{eqnarray}&&\!\!\!\!\!\!\!\!\!\!\!\!\!
\psi(u)\!\approx\! 1\!-\!\langle \tau\rangle u\!+\!\frac{u^2}{2}\int_0^{T_0}\!\!\! \psi(\tau)\tau^2 d\tau\!-\!\frac{(1\!-\!2R)^2 u^2\ln u}{\pi R}.\label{asuy}
\end{eqnarray}
We see that the first three terms in $\psi(u)$ are due to the normal diffusion events (that also determine $\langle \tau\rangle$). Long motions in infinite corridors contribute the last, logarithmic, term whose relevance depends on the crowding parameter that at $R$ close to one half obeys,
\begin{eqnarray}&&\!\!\!\!\!\!\!
\zeta\equiv \frac{A_2}{A_1}\sim (1-2R)^{-2},\ \ R\to 1/2. \label{crowding}
\end{eqnarray}
The logarithm in Eq.~(\ref{sml}) is not that large at realistic times ($u$ and time are inversely proportional) so that there can be situations where the normal diffusion $A_2$ term dominates the logarithmic term or is comparable with it. It was found in \cite{us} that $\zeta\simeq 9$ at $R=0.4$ so that $\zeta$ can be large at $R$ which is not very close to the finite-infinite horizon threshold. This has significant implications for the particle's dispersion considered below. In contrast, $A_1\propto R^{-2}$ behavior at small $R$ implies vanishing of the crowding parameter at small $R$ (strictly speaking this is a conjecture whose the full proof demands the study of small $R$ behavior of $A_2$. This is beyond the scope of this work that does not concentrate on $R\to 0$ limit). Similar considerations can be made for triangular \cite{bleher} or other lattices.

The structure of the asymptotic expansion in Eq.~(\ref{sml}) can be understood by considering the reference distribution $\psi_0(\tau)$ defined by,
\begin{eqnarray}&&\!\!\!\!\!\!\!\!\!\!\!\!\!\!\!\!\!\!
\psi_0(\tau)=\frac{2}{\tau^{3}}, \ \ \tau>1; \ \ \psi(\tau)=0,\ \ 0\leq\tau\leq 1.\label{model1}
\end{eqnarray}
For this function $\langle \tau\rangle=2$ and,
\begin{eqnarray}&&\!\!\!\!\!\!\!\!\!\!\!\!\!\!
\psi_0''\!=\!2\int_1^{\infty}\!\!\frac{\exp[-u\tau]d\tau}{\tau}\!=\!-2Ei(-u)
\!\sim\! -2\left(C\!+\!\ln u\right),
\end{eqnarray}
where $Ei(z)$ is the exponential integral and $C=0.577215$ is Euler's constant. Comparing this with $\psi_0''(u)=-2A_1\ln u-3A_1+2A_2$ we find that for this function Eq.~(\ref{sml}) holds with $A_1=1$ and $A_2=-C+1.5$. For this function $\psi_0(u)+2\ln u$ is an analytic function. The $u^{5/2}$ term in Eq.~(\ref{sml}) comes from the leading order, $\tau^{-7/2}$, correction to the cubic tail of $\psi(\tau)$ at large times \cite{bleher}.

The particle's displacement in time $t$ is given by the LW as (here the initial velocity is the velocity of the first step $\bm v_1$),
\begin{eqnarray}&&\!\!\!\!\!\!\!\!\!
\bm r(t)=\sum_{i=1}^{N(t)} \bm v_i\tau_i+\bm v_{N(t)+1}\tau^*,
\end{eqnarray}
where $N(t)$ is the number of renewals that occur in time $t$. We defined the backward recurrence time $\tau^*\equiv t-\sum_{i=1}^{N(t)}\tau_i$ that gives the time that elapsed since the last renewal. If $N(t)=0$ then $t=0$ is considered as the last renewal and the sums from one to $N(t)$ are defined as zero. We investigate the PDF $P(\bm r, t)$ of the walker's displacement $\bm r(t)$ in time $t$. The Fourier-Laplace transform of this function,
\begin{eqnarray}&&\!\!\!\!\!\!\!\!\!
 P(\bm k, u)=\int_0^{\infty} dt \int d\bm r\exp\left(-i\bm k\cdot\bm r-ut\right) P(\bm r, t),\label{mw}
\end{eqnarray}
obeys the Montroll-Weiss equation \cite{UchZol,Zaburdaev,2017},
\begin{eqnarray}&&\!\!\!\!\!\!\!\!\!\!\!\!\!\!
P(\bm k, u)=\left\langle\frac{1-\psi(u-i\bm k\cdot\bm v)}{u-i\bm k\cdot\bm v}\right\rangle\frac{1}{1-\left\langle \psi(u-i\bm k\cdot\bm v)\right\rangle},\label{basic}
\end{eqnarray}
where the angular brackets stand for averaging over the statistics of velocity and $\psi(u)$ is the Laplace transform of $\psi(\tau)$. We are interested in the asymptotic properties of the PDF and the moments of $\bm r(t)$ at large times that correspond to the small argument limit of the inverse Laplace transform.

\begin{figure*}
\includegraphics[width=8.5cm]{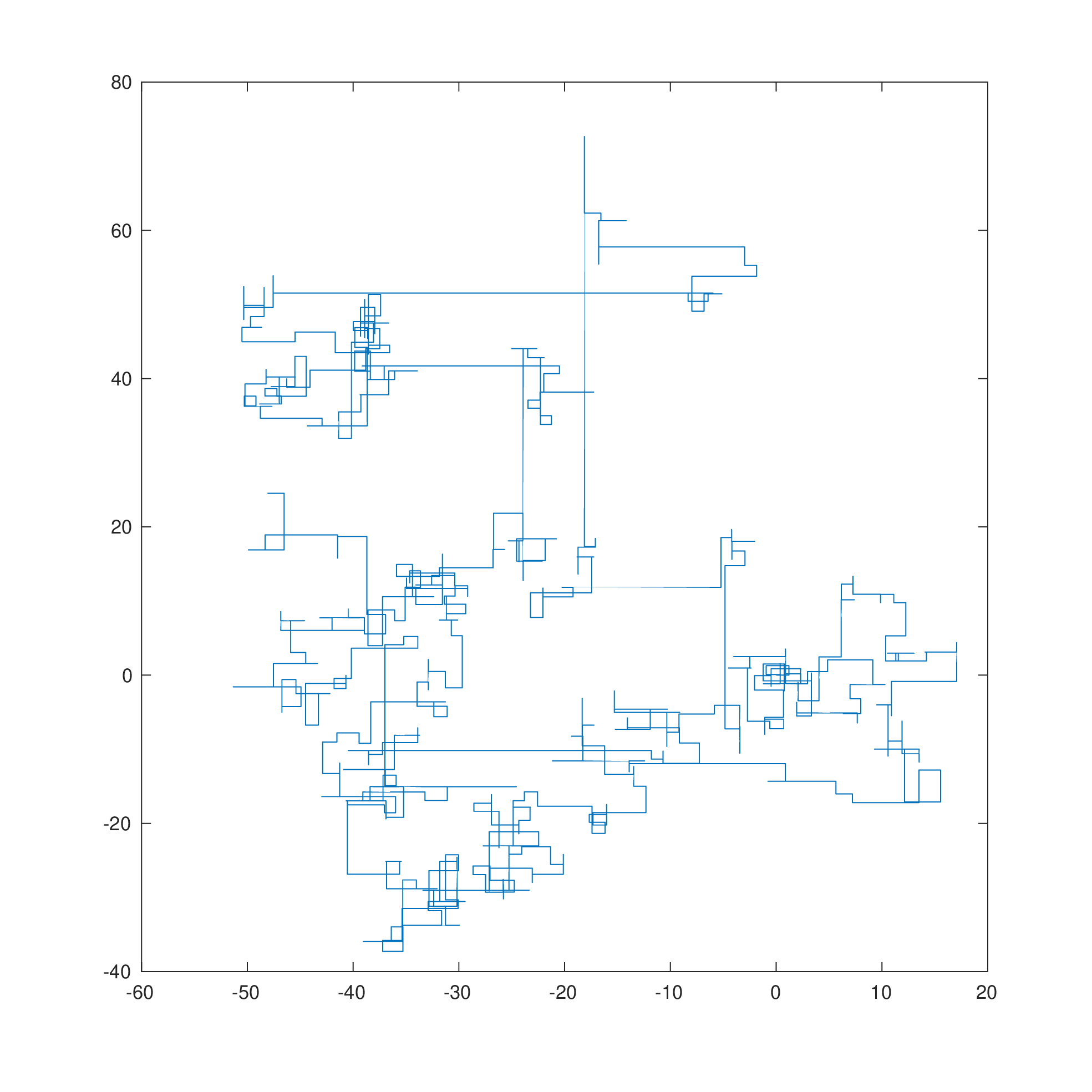}
\caption{Typical realization of the LW defined by Eqs.~(\ref{vela}) and (\ref{model1}). The particle propagates by diffusive-like motions interrupted by long ballistic flights.}
\label{realization}
\end{figure*}

The calculations below are done for general velocity statistics up to restrictions imposed in the beginning of this Section. As an example we will use the statistics given by Eq.~(\ref{vela}) where the coupling of motions in $x$ and $y$ directions is due to the common time resource that these motions split. Thus if $\rho(t', t)$ is the PDF that the during the time interval $t$ the particle moved parallel to $x$ axis during the time $t'$ then,
\begin{eqnarray}&&\!\!\!\!\!\!\!\!\!\!\!\!\!\!
P(x, y, t)=\int_0^t P(x, t') P(y, t-t')\rho(t', t) dt',
\end{eqnarray}
where $P(x, t)$ is the PDF of one-dimensional version of our walk and $x(0)=y(0)=0$. We will not use this representation for not limiting our consideration to Eq.~(\ref{vela}) however it might be of use in future studies. The distribution $\rho(t', t)$ can be found using the techniques of \cite{godreche}. A typical realization of the LW is provided in Fig. \ref{realization}.

\section{Diffusion coefficient at large times}\label{if}

In this Section we provide the formula for the displacement's dispersion that is valid at all times. This is given in terms of $\psi(u)$ and is an immediate consequence of the Montroll-Weiss equation. The evaluation at large times gives Eq.~(\ref{growth}).

\begin{figure*}
\includegraphics[width=8.5cm]{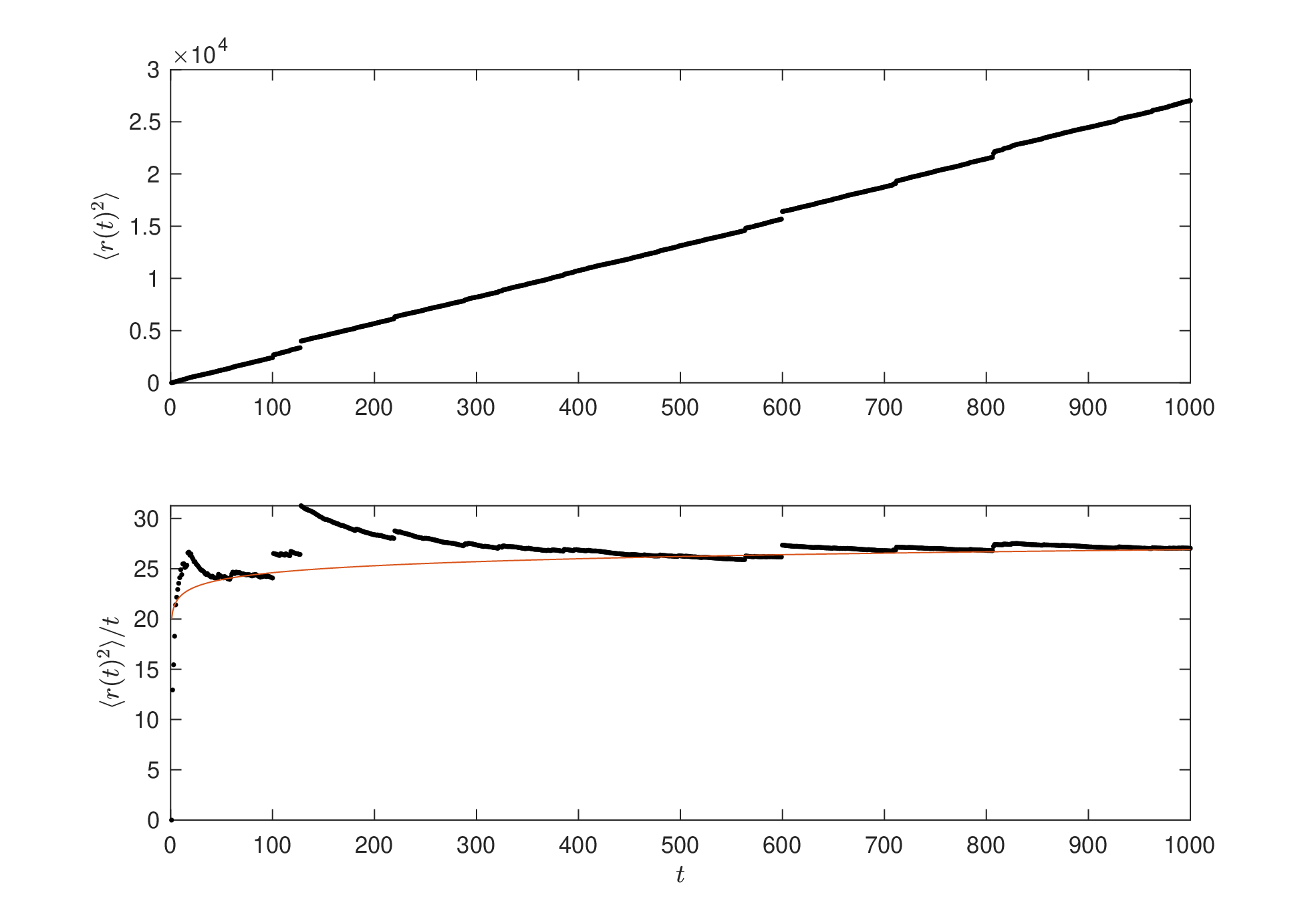}
\caption{The growth of dispersion for the LW defined by Eqs.~(\ref{vela}) and (\ref{model1}) with solid line providing the theoretical prediction. Rare long flights cause visible discontinuities despite averaging over $ 5 \times 10^8$ trajectories. }
\label{spersion}
\end{figure*}

The finite-time diagonal component of the diffusion matrix (diffusion coefficient in direction $i$), see e. g. \cite{det}, can be defined as \cite{bleher},
\begin{eqnarray}&& \!\!\!\!\!\!\!\!\!\!\!
D_{ii}(t)\equiv \frac{\langle r_i^2(t)\rangle}{2t},
\end{eqnarray}
where $i$ is an index of cartesian coordinates. We have,
\begin{eqnarray}&& \!\!\!\!\!\!\!\!\!\!\!\!\!\!\!\!\!\!
\langle r_i^2\rangle\!=\!-\frac{\partial^2}{\partial k_i^2}\left[\left\langle\frac{1\!-\!\psi(u\!-\!i\bm k\cdot\bm v)}{u\!-\!i\bm k\cdot\bm v}\right\rangle\frac{1}{1\!-\!\left\langle
\psi(u\!-\!i\bm k\cdot\bm v)\right\rangle}\right],
\end{eqnarray}
where the derivative is taken at $\bm k=0$. 
We find using that odd moments of $v_i$ vanish that \cite{2017},
\begin{eqnarray}&&\!\!\!\!\!\!\!\!
\langle r_i^2\rangle
=2\left\langle v_i^2\right\rangle\frac{u \psi'(u)-\psi(u)+1}{u^3[1-\psi(u)]}.\label{complete}
\end{eqnarray}
Inverse Laplace transform can be used for finding detailed time dependence of $\langle r_i^2(t)\rangle$ for a given $\psi(u)$. The long-time asymptotic form is universal and can be obtained from,
\begin{eqnarray}&&\!\!\!\!\!\!\!\!\!\!\!\!\!\!\!\!
\frac{u \psi'(u)\!+\!1\!-\!\psi(u)}{u^3[1\!-\!\psi(u)]}\!=\!\frac{\!-\!{\tilde A}_1\ln u\!-\!{\tilde A}_1\!+\!{\tilde A}_2\!+\!O(u^{1/2})}{u^2},
\end{eqnarray}
where ${\tilde A}_i=A_i/\langle \tau\rangle$ and we used Eq.~(\ref{sml}). Using that inverse Laplace transform of $-\ln u/u^2$ is $t\ln t-(1-C)t$ we have,
\begin{eqnarray}&&\!\!\!\!\!\!\!\!\!\!\!
\langle r_i^2(t)\rangle\!=\! \frac{2\left\langle v_i^2\right\rangle A_1t}{\langle \tau\rangle} \left(\ln t\!+\!\zeta\!-\!2\!+\!C\!+\!O\left(t^{-1/2}\right)\right). \label{growth01}
\end{eqnarray}
Summing over $i$, we find Eq.~(\ref{growth}) from the Introduction with $\zeta\equiv A_2/A_1$. We have for the case given by Eq.~(\ref{model1}),
\begin{eqnarray}&&
\frac{\langle r_i^2\rangle}{\left\langle v_i^2\right\rangle}\sim t\ln t-0.5t. \label{dasp}
\end{eqnarray}
The results of the simulations are provided in Fig. \ref{spersion}. The leading order term cannot describe the observations despite that the simulation was quite long, but including the correction linear in $t$ fits well, cf. \cite{cr1} and the next Section.

\section{Fast convergent CLT} \label{fast}

In this Section we consider the question of finding a factor $A_N$ for the rescaled sum $\sum_{i=1}^N \bm \xi_i/A_N$ of many independent identically distributed vector random variables $\bm \xi_i$ so that
the convergence to the Gaussian distribution is fastest. We assume that the distribution of $\bm \xi_i$ comes from the LW, $\bm \xi_i=\bm v_i\tau_i$.

It is usual to seek for the scaling factor so that the limiting distribution at $N\to \infty $ is Gaussian without considering the question of the convergence rate, see e. g. \cite{gned,bg}. The result that the distribution of $\sum_{i=1}^N \bm \Delta_i/\sqrt{N\ln N}$ at $N\to\infty$ is Gaussian belongs to this category \cite{bleher}. However in our case of logarithmically divergent $\langle \xi_i^2\rangle$ the convergence condition involves logarithmic factors that are never too large which poses the question of finding a rapidly convergent form.
The fast convergent CLT for the continuous time was found in \cite{us} and is considered in the next Section. Here we demonstrate that the Lambert scaling of \cite{us} is more intuitive in the discrete case. We also provide here the passage from discrete to continuous time cases that gives
the usual (not fast convergent) CLT, cf. \cite{UchZol,st}.

\subsection{Lambert scaling for discrete LW sum}

We consider the case of the LW $\bm \xi_i=\bm v_i\tau_i$ and study the distribution of,
\begin{eqnarray}&& \!\!\!\!\!\!\!\!\!\!\!\!\!\!
\bm Y_N=\frac{\sum_{i=1}^{N}\bm v_i\tau_i}{A_N}.\label{def}
\end{eqnarray}
We have for the characteristic function,
\begin{eqnarray}&& \!\!\!\!\!\!\!\!\!\!\!\!\!\!
P_N(\bm k)=\left\langle \exp[i\bm k \cdot \bm Y_N]\right\rangle=\left\langle \exp\left(\frac{i\bm k \cdot \bm v \tau}{A_N}\right)\right\rangle^N,
\end{eqnarray}
where we used the independence of the summands in $\bm Y_N$. Performing averaging over $\tau$,
\begin{eqnarray}&& \!\!\!\!\!\!\!\!\!\!\!\!\!\!
P_N(\bm k)=\left\langle \psi\left(u=\frac{\epsilon-i\bm k \cdot \bm v}{A_N}\right)\right\rangle^N,
\end{eqnarray}
where infinitesimal $\epsilon$ is introduced because the domain of definition of the Laplace transform $\psi(u)$ is given by complex $u$ with positive real part. We find performing averaging over $\bm v$ in the small argument expansion of $\psi(u)$,
\begin{eqnarray}&& \!\!\!\!\!\!\!\!\!\!\!\!\!
P_N(\bm k)\!\approx \!\left(1\!-\!\frac{A_1(\ln A_N\!+\!\zeta\!-\!\ln\left(\epsilon\!-\!i\bm k \!\cdot\! \bm v\right))\langle (\bm k\! \cdot \!\bm v)^2 \rangle}{A_N^2}\right)^N, \label{char}
\end{eqnarray}
where we neglected term of order $A_N^{-1/2}$, see Eq.~(\ref{sml}).
In the study of this formula below we will assume that $\ln\left(\epsilon\!-\!i\bm k \!\cdot\! \bm v\right)$ is of order one which is
valid in the study of the central part of the PDF of interest here.
We can have three different situations considered below.

{\it Nearly finite horizon, $\zeta\to\infty$.---} This is the case of $R$ close to one half, see Eq.~(\ref{crowding}). In this case we may use the usual scaling factor of the CLT, $A_N=\sqrt{N}$. We have then that at $N$ so large that $\sqrt{N}\gg 1$ however not so large that $\ln A_N\gtrsim \zeta$ the usual CLT holds.
We find neglecting in Eq.~(\ref{char}) the logarithmic terms in comparison with $\zeta=A_2/A_1$,
\begin{eqnarray}&& \!\!\!\!\!\!\!\!\!\!\!
P_N(\bm k)\!\approx \!\left(1\!-\!\frac{A_2\langle (\bm k\! \cdot \!\bm v)^2 \rangle}{N}\right)^N
=\exp\left(-A_2\langle (\bm k \cdot \bm v)^2 \rangle\right)
\nonumber\\&&\!\!\!\!\!\!\!\!\!\!\!\times
\left(1+O\left(\frac{A_2^2\langle (\bm k\! \cdot \!\bm v)^2 \rangle^2}{N}\right)\right).
\end{eqnarray}
We use here that $A_2$ is finite at $R\to 1/2$ and thus introduces no other parameter in the consideration. We observe that the (discrete) walker behaves as if it moved in the LG with finite horizon. Indeed, assuming that the asymptotic form of $A_2$ at $R\to 1/2$, given by Eq.~(\ref{asuy}) applies, we have that,
\begin{eqnarray}&& \!\!\!\!\!\!\!\!\!\!\!
\left\langle \exp\left[i\bm k \cdot \left(\sum_{i=1}^{N}\bm v_i\tau_i\right)\right]\right\rangle
=\left\langle \exp\left(i\sqrt{N}\bm k \cdot \bm Y_N\right)\right\rangle
\nonumber\\&&\!\!\!\!\!\!\!\!\!\!\!\!\!\!
\approx\exp\left(-\frac{N\langle (\bm k \cdot \bm v)^2 \rangle \int_0^{T_0}\! \psi(\tau)\tau^2 d\tau}{2}\right),
\end{eqnarray}
which inverse Fourier transform provides the central part of the PDF of $\sum_{i=1}^{N}\bm v_i\tau_i$. This is the usual CLT that describes normal diffusion obtained by neglecting flights longer than $T_0$. At $1/2-R\ll 1$ the motions in the infinite corridors are too rare for influencing the PDF appreciably, see Eq.~(\ref{crowding}). This result holds at $\ln N\ll 2\zeta$ where the factor of two must be kept because of exponential sensitivity on it in terms of $N$. This domain in $N$ can be the largest available in the simulations \cite{cr1}. The infinite horizon becomes relevant only at the largest $N$ when $\ln A_N$ eventually becomes comparable with $\zeta$. We consider first the opposite limit when $\ln A_N$ is much larger (and not comparable) than $\zeta$. That limit holds at $N\to \infty$ or at a finite large $N$ when $\zeta$ is moderate.

{\it The limit of $\zeta\lesssim 1$ or $N\to\infty$.---} When the scatterers' radius is well below the finite-infinite horizon threshold of one half, we have $\zeta\lesssim 1$ and the Gaussian PDF becomes applicable at $\ln A_N\gg 1$. This is also the case of $\zeta\gg 1$ at $N\to\infty$ which observation however would be obstructed by very large $N$ involved, see \cite{cr1} and the discussion in \cite{us}. We have in these cases at $\ln A_N\gg \max[1, \zeta]$,
\begin{eqnarray}&& \!\!\!\!\!\!\!\!\!\!\!
P_N(\bm k)\!\approx \!\left(1\!-\!\frac{A_1 \ln A_N\langle (\bm k\! \cdot \!\bm v)^2 \rangle}{A_N^2}\right)^N.
\end{eqnarray}
We fix $A_N$ by the condition $A_N^2=(N A_1/2)\ln A_N$ so that the formula gets closest to the Gaussian similarly to the case considered above (the prefactor of $A_1/2$ is used for having transparent correspondence with continuous time case below). The solution for $A_N$ can be written with the help of the lower of the two real branches of the Lambert function, $W_{-1}(x)$, as,
\begin{eqnarray}&& \!\!\!\!\!\!\!\!\!\!\!\!\!\!
A_N^2=\frac{N A_1}{4}\left|W_{-1}\left(-\frac{4}{N A_1}\right)\right|, \ \
\frac{A_N^2}{\ln A_N}=\frac{NA_1}{2}. \label{defA}
\end{eqnarray}
We observe that our assumption $\ln A_N\gg \max[1, \zeta]$ can be written as $|W_{-1}(-4/N A_1)|\gg \max[1, \zeta]$. We find,
\begin{eqnarray}&& \!\!\!\!\!\!\!\!\!\!\!
P_N(\bm k)\!\approx \!\left(1\!-\!\frac{2\langle (\bm k\! \cdot \!\bm v)^2 \rangle}{N}\right)^N=\exp\left(-2\langle (\bm k \cdot \bm v)^2 \rangle\right)
\nonumber\\&&\!\!\!\!\!\!\!\!\!\!\!\times
\left(1+O\left(\frac{\langle (\bm k\! \cdot \!\bm v)^2 \rangle^2}{N}\right)\right).
\end{eqnarray}
The condition $\left| W_{-1}(-4/NA_1)\right|\gg \max[1, \zeta]$ guarantees that, unless the numerator contains a (very) large numerical factor, the last line above can be dropped and we find Gaussian distribution of $\bm Y_N$.

\begin{figure*}
\includegraphics[width=8.5cm]{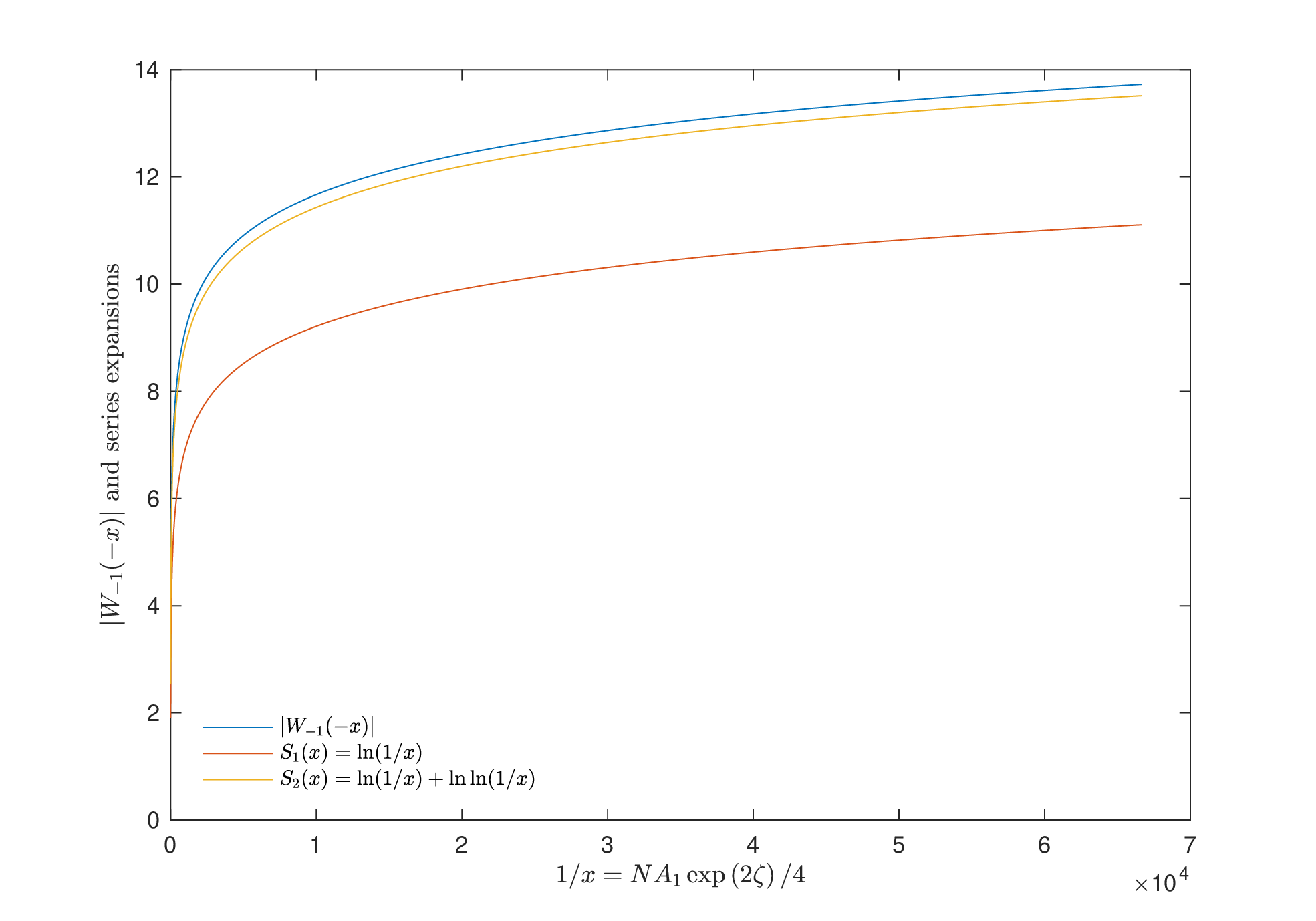}
\caption{Comparison of $\left|W_{-1}(-x)\right|$ with its series approximations $S_1$ and $S_2$ that contain one and two terms of the infinite series representation, respectively. It is seen that $S_2$ provides a good approximation when $\left|W_{-1}(-x)\right|\gg 1$ however $S_1$ is already quite reasonable. }
\label{LambertW_fig1}
\end{figure*}

The Lambert function provides a compact description of the result and can be reduced to elementary functions in the considered limit. We observe that the asymptotic series of $W_{-1}(-x)$ at $x\to 0$ implies that at $|W_{-1}(-4/N A_1)|\gg 1$,
\begin{eqnarray}&&\!\!\!\!\!\!\!\!\!\!\!\!\!\!
\left|W_{-1}\left(-\frac{4}{N A_1}\right)\right|\approx \ln  \left(\frac{NA_1}{4}\right) +\ln \ln  \left(\frac{N A_1}{4}\right), \label{alsn}
\end{eqnarray}
since the remainder of the series is of the order of $\ln |\ln x|/|\ln x|$ with $x=4/(NA_1)$ which is never much larger than one at relevant $N$, see Fig. \ref{LambertW_fig1}.

If we confine ourselves with accuracy of about twenty per cent, or if $N$ is (unrealistically) large then we may neglect the last term in Eq.~(\ref{alsn}). We find that under the condition $|W_{-1}(-4/NA_1)|\gg \max[1, \zeta]$
we can use $\left|W_{-1}\left(-4/N A_1\right)\right|\approx \ln (NA_1)$ and the condition
boils down to $\ln N\gg \max[1, \zeta]$. Here we dropped $\ln A_1$ term which is either of order one or diverges weaker than $\zeta$, see Eqs.~(\ref{as}) and (\ref{crowding}). We find $A_N^2\approx N A_1\ln N/4$, which is equivalent to the result of \cite{bleher}. We conclude that,
\begin{eqnarray}&&\!\!\!\!\!\!\!\!\!\!\!\!\!\!
\left\langle \exp\left[i\bm k\! \cdot\! \left(\sum_{i=1}^{N}\bm v_i\tau_i\right)\right]\right\rangle
\!=\!\left\langle e^{i\sqrt{(N A_1/4)\ln N}\bm k \cdot \bm Y_N}\right\rangle
\nonumber\\&&\!\!\!\!\!\!\!\!\!\!\!\!\!\!
\approx
\exp\left(-\frac{A_1  \langle (\bm k \cdot \bm v)^2 \rangle N \ln N}{2}\right),\label{dr}
\end{eqnarray}
which can be used for calculating the center of the PDF of the sum. The condition of validity of this result, $\ln\ln N/\ln N\ll 1$, observed in \cite{bleher}, is very stringent (unless $\simeq 0.2$ is considered to be much less than one), however inclusion of one more logarithm, as prescribed by Eq.~(\ref{alsn}), gives a Gaussian distribution under the much milder conditions. 

{\it General case of $\ln A_N+\zeta\gg 1$.---} This case includes the above two situations as limiting cases. Eq.~(\ref{char}) becomes,
\begin{eqnarray}&& \!\!\!\!\!\!\!\!\!\!\!
P_N(\bm k)\!\approx \!\left(1\!-\!\frac{A_1(\ln A_N\!+\!\zeta)\langle (\bm k\! \cdot \!\bm v)^2 \rangle}{A_N^2}\right)^N,
\end{eqnarray}
In this case we fix $A_N$ by the condition,
\begin{eqnarray}&& \!\!\!\!\!\!\!\!
\frac{2 A_N^2}{\ln A_N\!+\!\zeta}\!=\!N A_1, \ \ A_N^2\!=\!\frac{N A_1}{4}\left|W_{-1}\left(-\frac{4\exp\left(-2\zeta\right)}{N A_1}\right)\right|. \nonumber
\end{eqnarray}
The rest of the steps is similar.
Thus the distribution of,
\begin{eqnarray}&& \!\!\!\!\!\!\!\!\!\!\!\!\!\!
\bm Y_N=\frac{2\sum_{i=1}^{N}\bm v_i\tau_i}{\sqrt{N A_1|W_{-1}(-4\exp\left(-2\zeta\right)/N A_1)|}},\label{refine}
\end{eqnarray}
becomes a Gaussian, $N-$independent distribution,
\begin{eqnarray}&& \!\!\!\!\!\!\!\!\!\!\!\!\!\!
\left\langle \exp[i\bm k \cdot \bm Y_N]\right\rangle\approx \exp\left(-2\langle (\bm k \cdot \bm v)^2 \rangle\right), \label{ags}
\end{eqnarray}
at $N$ obeying $|W_{-1}(-4\exp\left(-2\zeta\right)/N A_1)|\gg 1$. We have,
\begin{eqnarray}&&\!\!\!\!\!\!\!\!\!\!\!\!\!\!
\left\langle \exp\left(i\bm k \cdot \left(\sum_{i=1}^{N}\bm v_i\tau_i\right)\right)\right\rangle
=\left\langle \exp\left(iA_N\bm k \cdot \bm Y_N\right)\right\rangle
\label{dr1}\\&&\!\!\!\!\!\!\!\!\!\!\!\!\!\!
\approx
\exp\left(-\frac{N A_1  \langle (\bm k\! \cdot\! \bm v)^2 \rangle |W_{-1}(-2\exp\left(-4\zeta\right)/N A_1)|}{2}\right).\nonumber
\end{eqnarray}
The reduction of the Lambert function to elementary functions is realized via,
\begin{eqnarray}&&\!\!\!\!\!\!\!\!\!\!\!\!\!\!
\left|W_{-1}\left(-\frac{4\exp\left(-2\zeta\right)}{NA_1}\right)\right|\approx \ln  \left(\frac{NA_1\exp\left(2\zeta\right)}{4}\right)
\nonumber\\&&\!\!\!\!\!\!\!\!\!\!\!\!\!\!
+\ln \ln  \left(\frac{NA_1\exp\left(2\zeta\right)}{4}\right),\label{lasoi}
\end{eqnarray}
where again we can neglect the last term if $N$ is extremely large or the required accuracy is not too high. We find then that the condition $|W_{-1}(-4\exp\left(-2\zeta\right)/N A_1)|\gg 1$ boils down to $\ln N+2\zeta\gg 1$ that interpolates between the conditions of the cases considered previously. We have $A_N^2\approx N A_1\left(\ln N+2\zeta\right)/4$ using which we can simplify Eq.~(\ref{refine}) by stating that the
PDF of,
\begin{eqnarray}&& \!\!\!\!\!\!\!\!\!\!\!\!\!\!
\bm Y_N=\frac{2\sum_{i=1}^{N}\bm v_i\tau_i}{\sqrt{NA_1\left(\ln N+2\zeta\right)}},\label{refine1}
\end{eqnarray}
is approximately Gaussian at $\ln N+2\zeta\gg 1$ as described by Eq.~(\ref{ags}).
This result reduces to that of \cite{bleher} at $N\to\infty$ however the practical relevance of the $\zeta$ correction can be most profound. For instance in the case of $R=0.4$ with $\zeta\simeq 9$ the value of $\ln N$ will never reach $2\zeta$ in the simulations performed presently. Thus the result of \cite{bleher} would not apply in practice, however Eq.~(\ref{refine1}) would. The corresponding form of Eq.~(\ref{dr1}) is,
\begin{eqnarray}&&\!\!\!\!\!\!\!\!\!\!\!\!\!\!
\left\langle \exp\left[i\bm k \cdot \left(\sum_{i=1}^{N}\bm v_i\tau_i\right)\right]\right\rangle\nonumber\\&&\!\!\!\!\!\!\!\!\!\!\!\!\!\!\approx
\exp\left(-\frac{A_1 N (\ln N+2\zeta) \langle (\bm k \cdot \bm v)^2 \rangle}{2}\right).
\end{eqnarray}
This equation can be used as the general fast convergent CLT that works at any $\zeta$, see also the next Section. The inclusion of the last term in Eq.~(\ref{lasoi}) does not change the described qualitative picture. 

Using the asymptotic results of \cite{bleher} on the low argument behavior of the characteristic function of one flight displacement $\bm \Delta$, the above results can be generalized to the LG case of $\bm \xi_i=\bm \Delta$ or other cases of $\bm \xi_i$ with logarithmically divergent dispersion.


\subsection{Passage to the continuous time}

The passage from the distribution of the discrete sum to continuous time is a standard problem, see e. g. \cite{bleher}. For moderate times this demands the techniques used in \cite{us}, considered in the next Section. However if we are interested only in the longest times described by $t\to\infty$ limit then the passage is straightforward. We can use that with probability one at $t\to\infty$ there is equality in distribution,
\begin{eqnarray}&&\!\!\!\!\!\!\!\!\!\!\!\!\!\!
\frac{\bm r(t)}{A_{t/\langle \tau\rangle}}=\frac{\sum_{i=1}^{N(t)}\bm v_i\tau_i}{A_{N(t)}},\label{rndsm}
\end{eqnarray}
where we used the law of large numbers $\lim_{t\to\infty}t/N(t)=\langle \tau\rangle$ for the number $N(t)$ of flights before time $t$, cf. \cite{2017}. We observed that the limiting distribution of rescaled $\bm r(t)$ coincides with that of the sum,
\begin{eqnarray}&&\!\!\!\!\!\!\!\!\!\!\!\!\!\!
\bm s(t)=\sum_{i=1}^{N(t)}\bm v_i\tau_i.\label{g2}
\end{eqnarray}
This is proved from the Montroll-Weiss equation using finiteness of $\langle\tau\rangle$, see \cite{2017} and Sec. \ref{sum}. The motion described by $\bm s(t)$ is the jump model described in Sec. \ref{open}. The significance of the equality of central parts of distributions of $\bm r(t)$ and $\bm s(t)$ is that at large times the displacement due to the last step of the walk can be neglected. Though this point could seem obvious, it is wrong for walks with infinite average $\tau$, see \cite{BarkaiPR}. We conclude that from Eq.~(\ref{rndsm}) and the results above that the limit in distribution,
\begin{eqnarray}&&\!\!\!\!\!\!\!\!\!\!\!\!\!\!
\lim_{t\to\infty}\frac{\bm r(t)}{(t\ln t)^{1/2}}=\bm \eta,
\end{eqnarray}
exists and $\bm \eta$ is a Gaussian random variable. This the LW counterpart of the LG's CLT of \cite{bleher}. Here fixing the width of the Gaussian distribution with the dispersion of $\bm r(t)$ is wrong. Indeed, the characteristic function $P(\bm k, t)$ of $\bm r(t)$
obeys at large times Eq.~(\ref{dr}) with $N=t/\langle \tau\rangle$,
\begin{eqnarray}&& \!\!\!\!\!\!\!\!\!\!\!
P\!\sim\! \exp\left(-\frac{A_1\langle v_i^2\rangle k^2 t\ln t }{2\langle \tau\rangle}\right)\!= \!\exp\left(-\frac{\langle r_i^2(t)\rangle_{\infty} k^2}{4}\right),\nonumber
\end{eqnarray}
where we used $\langle (\bm k\cdot\bm v)^2\rangle=\langle v_i^2\rangle k^2$ valid for any $i$, see Eq.~(\ref{vela}), and Eq.~(\ref{growth01}) at large times. The inverse Fourier transform gives,
\begin{eqnarray}&&\!\!\!\!\!\!\!\!\!\!\!\!\!\!
P(\bm r, t)\sim \prod_{i=1}^d 
\frac{1}{\sqrt{\pi \langle r_i^2(t)\rangle_{\infty}}}\exp\left(-\frac{r_i^2}{\langle r_i^2(t)\rangle_{\infty}}\right),
\end{eqnarray}
The denominator in the exponent of this Gaussian PDF is $\langle r_i^2(t)\rangle_{\infty}$ and not $2\langle r_i^2(t)\rangle_{\infty}$. Thus if we calculated the dispersion using this PDF we would find only half the true value. This doubling effect is known for the Lorentz gas \cite{dett14}.

The CLT in this form implies that considering the family of L\'{e}vy walk models having power-law tails $\psi(\tau)\sim \tau^{-1-\alpha}$ the value of $\alpha=2$ is singular. For $\alpha>2$ we would have the usual CLT without the doubling effect. Similarly taking the limit of $\alpha\to 2$ in the L\'{e}vy distribution holding at $1<\alpha<2$ would not produce the $\ln t$ factor. Thus $\alpha\neq 2$ and $\alpha=2$ cases are different.

The CLT at $\alpha=2$ has the degree of universality typical for the usual situation where the second moment of the velocity $\langle v_i v_k\rangle$ determines the distribution completely. In contrast, at $\alpha<2$ much more detailed information on the velocity statistics enters the counterpart of the CLT \cite{2017}.

\section{Refined continuous time CLT and doubling effect}\label{da}

In this Section we revisit the fast convergent CLT for the LW derived in \cite{us}. Our purpose to provide the simplified form that makes the result more transparent and sheds more light on the doubling effect.

We observe that the PDF of $N(t)/t$ is at large times strongly peaked at $1/\langle \tau\rangle$. Thus it is highly plausible that for the description of the central part of the PDF we can assume equality in distribution,
\begin{eqnarray}&&\!\!\!\!\!\!\!\!\!\!\!\!\!\!
\bm r(t)=\sum_{i=1}^{N=t/\langle \tau\rangle}\bm v_i\tau_i, \label{heu}
\end{eqnarray}
which difference from Eq.~(\ref{rndsm}) is that it is assumed to hold at finite large times. We find from Eq.~(\ref{dr1}),
\begin{eqnarray}&&\!\!\!\!\!\!\!\!\!\!\!\!\!\!
\left\langle \exp\left(i\bm k \cdot \bm r(t)\right)\right\rangle\approx
\left\langle \exp\left(i\bm k \cdot \left(\sum_{i=1}^{N=t/\langle \tau\rangle}\bm v_i\tau_i\right)\right)\right\rangle
\label{charac}\\&&\!\!\!\!\!\!\!\!\!\!\!\!\!\!
\approx\exp\left(-\frac{t A_1  \langle (\bm k\! \cdot\! \bm v)^2 \rangle |W_{-1}(-4\langle \tau\rangle\exp\left(-2\zeta\right)/t A_1)|}{2\langle \tau\rangle}\right).\nonumber
\end{eqnarray}
This formula was demonstrated to hold rigorously in \cite{us} who considered the velocity statistics given by Eq.~(\ref{vela}) where $\langle (\bm k\! \cdot\! \bm v)^2 \rangle=1/2$ for $V=1$. It was demonstrated in \cite{us} that for the two-dimensional LW considered here,
\begin{eqnarray}&&\!\!\!\!\!\!\!\!\!\!\!\!\!\!
P(\bm r, t)\approx \frac{1}{\pi\xi^2(t)}\exp\left(-\frac{r^2}{\xi^2(t)}\right),\label{gasq}
\end{eqnarray}
where,
\begin{eqnarray}&&\!\!\!\!\!\!\!\!\!\!\!\!\!\!
\xi^2(t)=\frac{A_1t}{\langle \tau\rangle}\left|W_{-1}\left(-\frac{4\langle \tau\rangle \exp(-2\zeta)}{A_1t}\right)\right|, \label{gas}
\end{eqnarray}
which is the inverse Fourier transform of Eq.~(\ref{charac}). This formula similarly to Eq.~(\ref{dr1}) holds at $\left|W_{-1}\left(-4\langle \tau\rangle \exp(-2\zeta)/(A_1t)\right)\right|\gg 1$. It was demonstrated numerically that Eqs.~(\ref{gasq})-(\ref{gas}) provide very good description of simulations of the LG starting from moderate times with $N(t)\sim 10^4$ that are readily attained numerically. This is consistent with the considerations of the previous Section since the Gaussianity condition $|W_{-1}(-4\exp\left(-2\zeta\right)/N A_1)|\gg 1$, for the value of $\zeta$ holding in the simulations of \cite{us}, holds starting from $N\sim 10^4$. 
In contrast, the CLT of \cite{bleher} could not be observed. The main reason for this can be seen using approximation of the Lambert function by one logarithmic term. That gives that instead of $\xi^2(t)$ in Eq.~(\ref{gas}) we can use,
\begin{eqnarray}&&\!\!\!\!\!\!\!\!\!\!\!\!\!\!
\xi^2(t)\!\approx\! \frac{A_1 t}{\langle \tau\rangle}\left(\ln \frac{t}{\langle \tau\rangle}\!+\!2\zeta\right),\label{cl}
\end{eqnarray}
provided that $\ln (t/\langle \tau\rangle)+2\zeta\gg 1$ and the next order correction can be disregarded. We call this formula mixed CLT since it combines normal and anomalous diffusions. The reduction to the separate cases of the normal and anomalous diffusions are obtained in complete similarity with the studies of the previous Section. We observe that the simulations of \cite{us} were done at $R\simeq 0.4$ and $\zeta\simeq 9$. In this case the regime of \cite{bleher} that would hold at $\ln (t/\langle \tau\rangle)\!\gg \!2\zeta$ is unreachable.

We observe that the dispersion $\xi^2(t)$ found using the Gaussian PDF above is half the full dispersion, given by Eq.~(\ref{growth}) only when the dispersion is dominated by the logarithmic term. The corrections to the logarithm destroy the doubling effect. In the regime of normal diffusion (not too large time, $R$ close to $1/2$ and $\zeta\to\infty$), the dispersion found from Eqs.~(\ref{gasq}) and (\ref{cl}) coincides with the full dispersion given by Eq.~(\ref{growth}) when only $\zeta$ term is kept.

The result given by Eqs.~(\ref{gasq})-(\ref{gas}) is rigorous \cite{us}. In contrast, our derivation of this result from the discrete case considered previously is heuristic as it relied on unproved Eq.~(\ref{heu}). This fits our purposes in the last two Sections which is simple demonstration of the origin of Eqs.~(\ref{gasq})-(\ref{gas}) and their reduction, given by Eq.~(\ref{cl}). In the next Section we perform the consistent study of the validity of Eq.~(\ref{heu}).

\section{Role of one step: dispersion} \label{sum}

The role of rare events in the formation of the dispersion can be demonstrated by comparing the dispersion considered in Sec. \ref{if} with that of the jump model introduced after Eq.~(\ref{g2}). The PDF $P_s(\bm s, t)$ of $\bm s(t)$ obeys the Montroll-Weiss equation (see e. g. \cite{2017} and references therein),
\begin{eqnarray}&& \!\!\!\!\!\!\!\!\!\!
P_s(\bm k, u)=\frac{1-\psi(u)}{u}\frac{1}{1-\left\langle \psi(u-i\bm k\cdot\bm v)\right\rangle}. \label{ref1}
\end{eqnarray}
This differs from Eq.~(\ref{basic}) by the structure of the first term. However both terms have identical behavior at small $u$ and $k$,
\begin{eqnarray}&& \!\!\!\!\!\!\!\!\!\!
\frac{1-\psi(u)}{u}\sim \left\langle\frac{1-\psi(u-i\bm k\cdot\bm v)}{u-i\bm k\cdot\bm v}\right\rangle\sim \langle\tau\rangle,\label{sa}
\end{eqnarray}
where we used Eq.~(\ref{sml}). Consequently the small $u$ and $k$ behaviors of $P(\bm k, u)$ and $P_s(\bm k, u)$ agree in the leading order. This implies the equality of long-time (small $u$) asymptotic behaviors of the central parts of the PDFs that describe the most probable events (Fourier transform at small $\bm k$ is close to normalization integral which is determined by the most probable events). This is the result that we used in Eq.~(\ref{rndsm}).

However the corrections to the leading order behavior provide for a difference of the LW and jumps models. The dispersion of $\bm s(t)$ obeys,
\begin{eqnarray}&& \!\!\!\!\!\!\!\!\!\!\!
\langle s_i^2\rangle\!=\!\frac{\psi(u)\!-\!1}{u}\! \frac{\partial^2}{\partial k_i^2}\!\frac{1}{1\!-\!\left\langle
\psi(u\!-\!i\bm k\cdot\bm v)\right\rangle}
\!=\!\frac{\langle v_i^2\rangle \psi''(u)}{u(1\!-\!\psi(u))} .
\end{eqnarray}
We observe that the dispersion depends on $\psi''$ in contrast with the dispersion of $\bm r(t)$ given by Eq.~(\ref{complete}). We have,
\begin{eqnarray}&& \!\!\!\!\!\!\!\!\!\!\!
\frac{\langle s_i^2\rangle}{\langle v_i^2\rangle}\sim\frac{-2{\tilde A}_1\ln u-3{\tilde A}_1+2{\tilde A}_2}{ u^2},
\end{eqnarray}
whose inverse Laplace transform gives (transform of $-\ln u/u^2$ is $t\ln t-(1-C)t$),
\begin{eqnarray}&& \!\!\!\!\!\!\!\!\!\!\!
\langle s_i^2(t)\rangle\sim \frac{2\langle v_i^2\rangle A_1 t}{\langle \tau\rangle}
\left(\ln t+\zeta-\frac{5}{2}+C\right).\label{se}
\end{eqnarray}
We see comparing with Eq.~(\ref{growth01}) that in the leading order at large times $\langle s_i^2(t)\rangle=\langle r_i^2(t)\rangle$. In the next order term however there is a finite difference between the dispersions. Thus the difference in only one step of the walk, which constitutes the difference of $\bm x$ and $\bm s$, gives a finite difference of dispersions at realistic time-scales where the linear term in $t$ is not negligible. Yet in the regime $\ln t+2\zeta\gg 1$ studied by the mixed CLT the dispersions will be similar. For instance in the case of $R=0.4$ and $\zeta=9$ the difference of Eqs.~(\ref{growth01}) and ~(\ref{se}) is negligible. In contrast, for the model given by Eq.~(\ref{model1}),
\begin{eqnarray}&&
\frac{\langle s_i^2\rangle}{\left\langle v_i^2\right\rangle}\sim  t\ln t-t,
\end{eqnarray}
which is significantly different from Eq.~(\ref{dasp}) unless $\ln \gg 1$. Finally we remark that the equality of the central parts of the PDFs of $\bm r$ and $\bm s$ in the asymptotic regime studied by the fast convergent CLT can be demonstrated fully by observing
that the derivation of \cite{us} of Eqs.~(\ref{gasq})-(\ref{gas}) does not involve corrections to Eq.~(\ref{sa}).

\section{Fourth moment and backward recurrence time}

In this Section we derive $\langle r^4(t)\rangle$ from the backward recurrence time $B_t$. This time is the time interval that passed since the last collision until the current moment $t$. Thus if the particle moved without any collisions then $B_t=t$. Otherwise $B_t=t-\sum_{i=1}^{N(t)}\tau_i$ where $N(t)$ is the number of collisions that occurred during the evolution time $t$. The statistics of $B_t$ is derived in \cite{godreche} for $\psi(\tau)$ whose tail is $\tau^{-1-\theta}$ with $\theta<2$.
These statistics are compactly reproduced here for studying the case of $\theta=2$ not considered in \cite{godreche}.

We observe that the average backward recurrence time determines the dispersion's rate of change. We have using $\bm r(t)=\sum_{i=1}^{N(t)} \bm v_i\tau_i+\bm v_{N(t)+1}(t-\sum_{i=1}^{N(t)}\tau_i)$ and $\bm r(t)=\bm v_1t$ for $N(t)=0$ that,
\begin{eqnarray}&&\!\!\!\!\!\!\!\!\!
\frac{d\langle r^2(t)\rangle}{dt}\!=\!2\langle \bm r(t)\!\cdot\!\bm v(t)\rangle\!= \!2\langle v^2\rangle\left( t \int_t^{\infty}\psi(\tau)d\tau
\right.\nonumber\\&&\!\!\!\!\!\!\!\!\!\left.
+\sum_{N=1}^{\infty} \left \langle \left(t\!-\!\sum_{i=1}^N\tau_i\right)
\theta\left(t\!-\!\sum_{i=1}^N\tau_i\right)\theta\left(\sum_{i=1}^{N+1}\tau_i\!-\!t\right) \right\rangle\right)\nonumber\\&&\!\!\!\!\!\!\!\!\!
\!=\! 2\langle v^2\rangle\langle B_t\rangle, \label{rec}
\end{eqnarray}
where $\theta(x)$ is the step function. We decomposed the average into the sum of contributions of mutually exclusive events characterized by different $N(t)$.
The product of the step functions guarantees that $N$ collisions occurred before time $t$.
The first term describes the contribution
of $N(t)=0$. It is proportional to $2A_1/t$ at large times and is negligible. The counterpart of Eq.~(\ref{rec}) would also be useful for the LG itself where correlations of velocities of different flights decay fast with the number of collisions between them \cite{bleher}. Similar equation for $\langle r^4(t)\rangle$ is considered later.

We derive the statistics of $B_t$ by considering its characteristic function in imaginary argument $f(\beta, t)=\langle \exp(-\beta B_t)\rangle$ which is the Laplace transform of the PDF of $B_t$ (the variable $u$ is reserved for the Laplace transform in $t$ below). This
function depends on $t$ as the variable that defines $B_t$. We have decomposing the average into the sum of contributions of the mutually exclusive events with different $N$ that,
\begin{eqnarray}&&\!\!\!\!\!\!\!\!
f(\beta, t)\!=\!e^{-\beta t}\int_t^{\infty}\!\!\! \psi(\tau)d\tau\!+\!
\sum_{N=1}^{\infty} \left\langle \exp\left(-\beta\left(t\!-\!\sum_{i=1}^N\tau_i\right)\right)
\right.\nonumber\\&&\!\!\!\!\!\!\!\!\left.
\theta\left(t-\sum_{i=1}^N\tau_i\right)\theta\left(\sum_{i=1}^{N+1}\tau_i-t\right) \right\rangle.
\end{eqnarray}
We perform Laplace transform in $t$ variable,
\begin{eqnarray}&& \!\!\!\!\!\!\!\!\!\!\!
f(\beta, u)\!=\!\frac{1\!-\!\psi(u\!+\!\beta)}{u\!+\!\beta}\!+\!
\sum_{N=1}^{\infty} \left\langle \int_{\sum_{i=1}^N\tau_i}^{\sum_{i=1}^{N+1}\tau_i}\exp\left(-\left(u\!+\!\beta\right)t
\right.\right.\nonumber\\&&\!\!\!\!\!\!\!\!\!\!\!\left.\left.
+\beta\sum_{i=1}^N\tau_i\right) dt\right\rangle,
\end{eqnarray}
where here and below $u$ is the Laplace transform variable. Taking the integral over $t$, 
averaging
over independent variables $\tau_i$ and summing the geometric series \cite{godreche},
\begin{eqnarray}&& \!\!\!\!\!\!\!\!\!\!\!
f(\beta, u)\!=\!\frac{1-\psi(u+\beta)}{(u+\beta)(1-\psi(u))}.
\end{eqnarray}
This compact formula contains all the statistics of $B_t$. We have for the average,
\begin{eqnarray}&& \!\!\!\!\!\!\!\!\!\!\!
\int_0^{\infty}\!\!\! \langle B_t\rangle e^{-u t}dt\!=\!-\partial_{\beta}f(0, u)\!=\!\frac{1}{1\!-\!\psi(u)}\frac{d}{du}\frac{\psi(u)\!-\!1}{u}.\label{fa}
\end{eqnarray}
This formula reproduces Eq.~(\ref{complete}) as seen by performing Laplace transform of Eq.~(\ref{rec}),
\begin{eqnarray}&& \!\!\!\!\!\!\!\!\!\!\!
\langle r^2(u)\rangle = \frac{2\langle v^2\rangle\langle B_t(u)\rangle}{u}
=\frac{2\langle v^2\rangle}{u(1\!-\!\psi(u))}\frac{d}{du}\frac{\psi(u)\!-\!1}{u}
\nonumber\\&&\!\!\!\!\!\!\!\!\!\!\!
=2\langle v^2\rangle \frac{u\psi'(u)\!-\!\psi(u)+1}{u^3(1\!-\!\psi(u))}.
\end{eqnarray}
We consider deriving a similar representation for the fourth order moment. We have from $d\langle r^4(t)\rangle/dt=4\langle r^2(t)\bm r(t)\cdot\bm v(t)\rangle$ that,
\begin{eqnarray}&&\!\!\!\!\!\!\!\!\!\!\!\!\!\!
\frac{1}{4}\frac{d\langle r^4(t)\rangle}{dt}\!=\! \langle v^4\rangle t^3\!+\!\sum_{N=1}^{\infty}\left \langle\theta\left(\sum_{i=1}^{N+1}\tau_i-t\right)
\right.\nonumber\\&&\!\!\!\!\!\!\!\!\!\!\!\left.\theta\left(t-\sum_{i=1}^N\tau_i\right)
\left(\sum_{i=1}^N \bm v_i\tau_i\!+\!\left(t\!-\!\sum_{i=1}^N\tau_i\right)\bm v_{N+1}\right)^2
\right.\nonumber\\&&\!\!\!\!\!\!\!\!\!\!\!\left.
\left(\sum_{i=1}^N \bm v_i\cdot\bm v_{N+1}\tau_i+\left(t-\sum_{i=1}^N\tau_i\right) v^2_{N+1}\right) \right\rangle.
\end{eqnarray}
We perform averaging over independent $\bm v_i$ using $F(-\bm v)=F(\bm v)$ and introducing $\langle v_{\alpha}v_{\beta}\rangle=T_{\alpha\beta}$ and $\kappa=2 tr T^2+\langle v^2\rangle^2$. We find that,
\begin{eqnarray}&&\!\!\!\!\!\!\!\!\!\!\!\!\!\!
\frac{1}{4}\frac{d\langle r^4(t)\rangle}{dt}= \langle v^4\rangle \langle B_t^3\rangle+\kappa \sum_{N=1}^{\infty}N\left \langle
\left(t-\sum_{i=1}^N\tau_k\right)\tau_1^2
\right.\nonumber\\&&\!\!\!\!\!\!\!\!\!\!\!\left.
\theta\left(t\!-\!\sum_{i=1}^N\tau_i\right)\theta\left(\sum_{i=1}^{N+1}\tau_i\!-\!t\right)\right\rangle,
\end{eqnarray}
where we observed that different terms in the sum over $\tau_i^2$ are identical.
We take another time derivative,
\begin{eqnarray}&&\!\!\!\!\!\!\!\!\!\!\!
\frac{1}{4}\frac{d^2\langle r^4(t)\rangle}{dt^2}= \langle v^4\rangle \frac{d\langle B_t^3\rangle}{dt}+\kappa \sum_{N=1}^{\infty}N\left \langle
\theta\left(t\!-\!\sum_{i=1}^N\tau_i\right)\tau_1^2
\right.\nonumber\\&&\!\!\!\!\!\!\!\!\!\!\!\left.
\theta\left(\sum_{i=1}^{N+1}\tau_i\!-\!t\right)\right\rangle
-\kappa \sum_{N=1}^{\infty}N\left \langle
\tau_1^2\int_0^t\delta\left(\sum_{i=1}^N\tau_i-t'\right)
\right.\nonumber\\&&\!\!\!\!\!\!\!\!\!\!\!\left.
\left(t-t'\right) \psi\left(t-t'\right)dt'\right\rangle,
\end{eqnarray}
where we averaged over $\tau_{N+1}$ and used $\theta\left(t\!-\!\sum_{i=1}^N\tau_i\right)=\int_0^t \delta\left(\sum_{i=1}^N\tau_i-t'\right)dt'$.
Performing Laplace transform over $t$ and using the small times behavior $\langle r^4(t)\rangle \propto t^4$ and $B_t\propto t$ we find,
\begin{eqnarray}&&\!\!\!\!\!\!\!\!\!\!\!\!\!\!\!\!
\langle r^4\rangle\!=\!\frac{4\langle v^4\rangle\langle B_t^3(u)\rangle}{u}
\!+\! \frac{4\kappa \psi''\left(1\!-\!\psi\!+\!u\psi'\right) }{u^3}\!\sum_{N=1}^{\infty}\! N\psi^{N-1},\label{fourth}
\end{eqnarray}
where we introduced $\langle B_t^3(u)\rangle\equiv \int_0^{\infty} \langle B_t^3\rangle e^{-u t}dt$. Here and below equations that give moments of the coordinate as a function of $u$ must be understood as the Laplace transform of the corresponding moment. Thus $\langle r^4\rangle$ in the equation above is the Laplace transform of $\langle r^4(t)\rangle$. We have similarly to Eq.~(\ref{fa}),
\begin{eqnarray}&& \!\!\!\!\!\!\!\!\!\!\!\!\!\!
\langle B_t^3(u)\rangle\!=\!-\partial_{\beta}^3f(0, u)\!=\!\frac{1}{1\!-\!\psi(u)}\frac{d^3}{du^3}\frac{\psi(u)\!-\!1}{u}.\nonumber
\end{eqnarray}
We obtain summing the series in Eq.~(\ref{fourth}),
\begin{eqnarray}&&\!\!\!\!\!\!\!\!\!\!\!\!\!\!
\langle r^4\rangle\!=\! \frac{4\langle v^4\rangle \langle B_t^3(u)\rangle}{u}\!+\! \frac{4\kappa \psi'\psi''\left(1\!-\!\psi\!+\!u\psi'\right) }{u^3(1-\psi)^2}.\label{fr}
\end{eqnarray}
We observe that at small $u$ the RHS is dominated by the $\langle B_t^3\rangle$ term. The leading order behavior of this term at small $u$ is found using Eq.~(\ref{sml}). We obtain,
\begin{eqnarray}&& \!\!\!\!\!\!\!\!\!\!\!\!\!\!
\frac{1}{1\!-\!\psi(u)}\frac{d^3}{du^3}\frac{\psi(u)\!-\!1}{u}\simeq \frac{A_1}{\langle \tau\rangle u^3}.
\end{eqnarray}
We find that at large times,
\begin{eqnarray}&& \!\!\!\!\!\!\!\!\!\!\!\!\!\!
\int_0^{\infty}\!\!\! \langle B_t^3\rangle e^{-u t}dt\simeq \frac{A_1}{\langle \tau\rangle u^3},\ \ \langle B_t^3\rangle \simeq \frac{A_1t^2}{2\langle \tau\rangle},
\end{eqnarray}
where the corrections are smaller by a power of $t$. We observe that $B_t$ is very different from what it would be for the fast decaying $\psi(t)$ where the moments of $B_t$ would be time-independent quantities proportional to powers of $\langle\tau\rangle$, cf. \cite{godreche}. The leading order behavior of the last term in Eq.~(\ref{fr}) at small $u$ is,
\begin{eqnarray}&&\!\!\!\!\!\!\!\!\!\!\!\!\!\!
\frac{\psi'\psi''\left(1\!-\!\psi\!+\!u\psi'\right) }{u^3(1-\psi)^2}\propto \frac{\ln^2 u}{u^3},
\end{eqnarray}
which is much smaller than $\langle B_t^3(u)\rangle/u\propto u^{-4}$ at large times. Thus,
\begin{eqnarray}&&\!\!\!\!\!\!\!\!\!\!\!\!\!\!
\langle r^4(t)\rangle\simeq \frac{2A_1\langle v^4\rangle t^3}{3\langle \tau\rangle}+O(t^2),\label{fourths}
\end{eqnarray}
where the $t^2$ correction contains also logarithmic factors. We conclude that both the second and the fourth moments of the particle's coordinate are determined at large times by the statistics of the backward recurrence time. This indicates that the study of statistics of $B_t$ might also be of interest for the LG.

We observe that the fourth moment grows much faster than the square of the fourth moment which is often referred as intermittency. This property holds because for most of the steps the particle advances a little and then it advances a lot by a long flight in the infinite corridor. These long flights form the far tail of $P(\bm r, t)$ derived in the next Section.

\section{Infinite density} \label{ifns}

In this Section we demonstrate that there is the finite scaling limit defining infinite density $I(\bm v)$,
\begin{eqnarray}&&\!\!\!\!\!\!\!\!\!\!\!\!\!\!
I(\bm v)=\lim_{t\to\infty}t^{d+1}P(t \bm v, t),\ \ \bm v\neq 0. \label{inf}
\end{eqnarray}
We demonstrate that as far as the PDF's tail is considered we can take the limit $\alpha\to 2$ from below of the infinite density tail in the L\'{e}vy walk with $\psi(\tau)\sim \tau^{-1-\alpha}$. However a certain change of prefactors is needed for getting finite limit: setting $\alpha=2$ in the infinite density tail obtained for $1<\alpha<2$ in \cite{BarkaiPhysRev,2017} gives ill-defined expressions.

We observe that (we switch to the argument $\bm w$ so that there is no confusion with $\bm v$ over which there is averaging below),
\begin{eqnarray}&&\!\!\!\!\!\!\!\!\!\!\!\!\!\!
P(t \bm w, t)=\int\frac{d\bm k'}{(2\pi)^d}\frac{du'}{2\pi i}\exp\left[it\bm k'\cdot\bm w+u't\right]{\tilde P}(\bm k', u')\nonumber\\&&\!\!\!\!\!\!\!\!\!\!\!\!\!\!=\frac{1}{t^{d+1}}\int\frac{d\bm k}{(2\pi)^d}\frac{du}{2\pi i}\exp\left[i\bm k\cdot\bm w+u\right]P\left(\frac{\bm k}{t}, \frac{u}{t}\right).\label{transform}
\end{eqnarray}
We consider the asymptotic form of $P\left(\bm k/t, u/t\right)$ at large $t$ and fixed $\bm k$, $u$ using the Montroll-Weiss equation given by Eq.~(\ref{basic}). We observe that,
\begin{eqnarray}&& \!\!\!\!\!\!\!\!\!\!\!\!\!\!
\left\langle\frac{1\!-\!\psi(ut^{-1}\!-\!i\bm k\cdot\bm vt^{-1})}{ut^{-1}\!-\!i\bm k\cdot\bm vt^{-1}}\right\rangle\!\sim\!\langle \tau\rangle\!+\!A_1\left\langle \left(
\frac{u}{t}\!-\!\frac{i\bm k\cdot\bm v}{t}\right)
\right.\nonumber\\&&\!\!\!\!\!\!\!\!\!\!\!\!\!\!\left.
\ln\left(\frac{u}{t}\!-\!\frac{i\bm k\cdot\bm v}{t}\right)\right\rangle-\frac{A_2 u}{t},\label{t1}
\end{eqnarray}
with corrections of order of $1/t$. Similarly (${\tilde A}_i=A_i/\langle\tau\rangle$), 
\begin{eqnarray}&& \!\!\!\!\!\!\!\!\!\!\!\!\!\! \left[1\!-\!\left\langle \psi\left(\frac{u}{t}\!-\!\frac{i\bm k\cdot\bm v}{t}\right)\right\rangle\right]^{-1}\!
\sim\!\frac{t}{\langle \tau\rangle u}\!+\frac{{\tilde A}_2}{\langle \tau\rangle}\left(1-\frac{k^2\langle v_i^2\rangle}{u^2}\right)
\nonumber\\&&\!\!\!\!\!\!\!\!\!\!\!\!\!\!
-\frac{{\tilde A}_1}{\langle \tau\rangle}\left\langle \left(1\!-\!\frac{i\bm k\cdot\bm v}{u}\right)^2\ln \left(\frac{u}{t}\!-\!\frac{i\bm k\cdot\bm v}{t}\right) \right\rangle.\label{t2}
\end{eqnarray}
We find from the Montroll-Weiss equation multiplying Eqs.~(\ref{t1}) and (\ref{t2}) that,
\begin{eqnarray}
&&
P\left(\frac{\bm k}{t}, \frac{u}{t}\right)\!\sim {\tilde A}_1\left\langle \left(\frac{i\bm k\cdot\bm v}{u}+\frac{(\bm k\cdot\bm v)^2}{u^2}\right)\ln \left(1\!-\!\frac{i\bm k\cdot\bm v}{u}\right) \right\rangle
\nonumber\\&&+\frac{t}{ u}
+{\tilde A}_2\left(1-\frac{k^2\langle v_i^2\rangle}{u^2}\right)
+\frac{{\tilde A}_1k^2\langle v_i^2\rangle\ln (u/t)}{u^2}.\label{s1}
\end{eqnarray}
The inverse Fourier transform of the last line involves only terms ($\delta-$function and its laplacian) that vanish at $\bm x\neq 0$. We find from Eq.~(\ref{transform}) at $\bm w\neq 0$,
\begin{eqnarray}&&\!\!\!\!\!\!\!\!\!\!\!\!\!\!
\lim_{t\to\infty}t^{d+1} P(t \bm w, t)={\tilde A}_1\int\frac{d\bm k}{(2\pi)^d}\frac{du}{2\pi i}\exp\left[i\bm k\cdot\bm w+u\right]
\nonumber\\&&\!\!\!\!\!\!\!\!\!\!\!\!\!\!\times
\left\langle \left(\frac{i\bm k\cdot\bm v}{u}+\frac{(\bm k\cdot\bm v)^2}{u^2}\right)\ln \left(1\!-\!\frac{i\bm k\cdot\bm v}{u}\right) \right\rangle.\label{limit}
\end{eqnarray}
This proves that the limit for $I(\bm w)$ in Eq.~(\ref{inf}) is finite.
For finding the integral in Eq.~(\ref{limit}) we use the series,
\begin{eqnarray}&&\!\!\!\!\!\!\!\!\!\!\!\!\!\!
(x-x^2)\ln(1-x)=\sum_{n=3}^{\infty}\frac{x^{n}}{(n-2)(n-1)}-x^2.
\end{eqnarray}
We find,
\begin{eqnarray}&&\!\!\!\!\!\!\!\!\!\!\!\!\!\!
\left\langle\left(\frac{i\bm k\cdot\bm v}{u}+\frac{(\bm k\cdot\bm v)^2}{u^2}\right)\ln \left(1\!-\!\frac{i\bm k\cdot\bm v}{u}\right) \right\rangle
\nonumber\\&&\!\!\!\!\!\!\!\!\!\!\!\!\!\!
=\sum_{n=2}^{\infty} \left\langle  \frac{(-1)^n(\bm k\cdot\bm v)^{2n}}{u^{2n}(2n-2)(2n-1)} \right\rangle+\frac{k^2\langle v_i^2\rangle}{u^2}.
\end{eqnarray}
We find integrating over $u$ (we discard the last term using $\bm w\neq 0$),
\begin{eqnarray}&&\!\!\!\!\!\!\!\!\!\!\!\!\!\!
I(\bm w)={\tilde A}_1\int\frac{d\bm k}{(2\pi)^d}\exp\left[i\bm k\cdot\bm w\right]\left\langle G(\bm k\cdot\bm v) \right\rangle
\nonumber\\&&\!\!\!\!\!\!\!\!\!\!\!\!\!\!
G(y)=\sum_{n=2}^{\infty}  \frac{(-1)^n y^{2n}}{(2n-1)!(2n-2)(2n-1)}.\label{tr}
\end{eqnarray}
This can be written as $I(\bm w)=\lim_{\alpha\to 2}|\Gamma(1-\alpha)|I_{\alpha}(\bm w)$ where,
\begin{eqnarray}&&\!\!\!\!\!\!\!\!\!\!\!\!\!
I_{\alpha}(\bm w)={\tilde A}_1\int\frac{d\bm k}{(2\pi)^d}\!\frac{\exp\left[i\bm k\!\cdot\!\bm w\right]\left\langle G_{\alpha}(\bm k\!\cdot\!\bm v)\right\rangle}{|\Gamma(1-\alpha)|},\nonumber\\&&\!\!\!\!\!\!\!\!\!\!\!\!\!
G_{\alpha}(y)=\sum_{n=1}^{\infty}\frac{(-1)^n y^{2n}}{(2n-1)!(2n-\alpha)(2n+1-\alpha)},
\end{eqnarray}
where we use $\bm w\neq 0$ for disregarding the $n=1$ term in the series. The function $I_{\alpha}(\bm w)$ is the infinite density of L\'{e}vy walk with the tail $\psi(\tau)\sim \tau^{-1-\alpha}$ that was found in \cite{2017}. We find switching $\bm w$ with $\bm v$,
\begin{eqnarray}&&\!\!\!\!\!\!\!\!\!\!\!\!\!\!
I(\bm v)\!=\frac{{\tilde A}_1}{v^{d-1}}\int_{v'>v} F(v'{\hat v})v'^{d-1} \left[2\frac{v'^{2}}{v^3}-\frac{v'}{v^{2}}\right]d v',\label{infdensity}
\end{eqnarray}
where the dependence in the velocity's PDF $F(\bm v)$ on the magnitude $v$ and direction ${\hat v}$ is separated. This result uses $F(\bm v)=F(-\bm v)$. We have $I(\bm v)\equiv 0$ for $v>nv_0$ since $F(\bm v)$ vanishes there.

We consider the infinite density for the velocity distribution given by Eq.~(\ref{vela}).
We find using $d=2$ that,
\begin{eqnarray}&&\!\!\!\!\!\!\!\!\!\!\!\!\!\!
I(\bm v)\!=\!\frac{{\tilde A}_1\theta(V-v)}{4} \left(\delta(v_x)+\delta(v_y)\right)\left[2\frac{V^{2}}{v^3}-\frac{V}{v^{2}}\right],
\end{eqnarray}
where $\theta(x)$ is the step function and we do not set $V=1$ for transparency. We conclude from Eq.~(\ref{inf}) that the tail of the distribution of the displacement obeys,
\begin{eqnarray}&&\!\!\!\!\!\!\!\!\!\!\!\!\!\!
P_{tail}(\bm r, t)\!\simeq \!\frac{{\tilde A}_1\theta(Vt\!-\!r)}{4} \left(\delta(r_1)\!+\!\delta(r_2)\right)\left(2\frac{V^{2}t}{r^3}\!-\!\frac{V}{r^{2}}\right),
\label{soltuioi}
\end{eqnarray}
The consistency of this tail can be seen be considering the probability of the maximal possible displacement $Vt$. We have,
\begin{eqnarray}&&\!\!\!\!\!\!\!\!\!\!\!\!\!\!
\int_{Vt-\epsilon<r<Vt} P_{tail}(\bm r, t) d\bm r\simeq
{\tilde A}_1\int_{Vt-\epsilon}^{Vt} dr_1  \left(2\frac{V^{2}t}{r_1^3}\!-\!\frac{V}{r_1^{2}}\right)\nonumber\\&&\!\!\!\!\!\!\!\!\!\!\!\!\!\!
\approx \frac{{\tilde A}_1\epsilon}{Vt^2}.
\end{eqnarray}
This quantity can also be found directly as,
\begin{eqnarray}&&\!\!\!\!\!\!\!\!\!\!\!\!\!\!
\int_{t-\epsilon/V}^t\psi(\tau)d\tau\approx \frac{2A_1 \epsilon}{V t^2}
\end{eqnarray}
This gives for the contribution of the infinite tail into the dispersion,
\begin{eqnarray}&&\!\!\!\!\!\!\!\!\!\!
\langle r^2(t)\rangle_{tail}\!\simeq \!\!\int_{r>r_c}\!\!\!\!\! r^2 P(\bm r, t)d\bm r\simeq \frac{{\tilde A}_1}{2}\!\int_{r_c}^{Vt} \!\!\!\! r_1^2\left[2\frac{V^{2}t}{r_1^3}\!-\!\frac{V}{r_1^{2}}\right]dr_1\nonumber\\&&
\!\!\!\!\!\!\!\!\!\!
\simeq \frac{{\tilde A}_1}{2}\left[2V^{2}t\ln (Vt)-V^2t\right],
\end{eqnarray}
where $r_c$ is the lower cutoff so that $P(\bm r, t)\approx P_{tail}(\bm r, t)$ for $r>r_c$.
Finally using $\langle v^2\rangle=V^2$ we find that in the leading order,
\begin{eqnarray}&&\!\!\!\!\!\!\!\!\!\!\!\!\!\!
\langle r^2(t)\rangle_{tail}= \frac{\langle v^2\rangle A_1 t\ln t}{\langle\tau\rangle}, \label{sdf}
\end{eqnarray}
which comparison with Eq.~(\ref{growth}) proves that the infinite density tail contributes one half of the leading order $t\ln t$ term in the dispersion of $\bm r(t)$. However, for the corrections, that are relevant, the doubling effect is no longer true. This is in accord with the results of Sec. \ref{da}.

The leading order term in $\langle r^2(t)\rangle_{tail}$ is not contributed by the last term in Eq.~(\ref{soltuioi}). This does not tell that this term is negligible since its contribution in the higher order moments, considered in the next Section, is of the same order as the first term.

We return to the general $F(\bm v)$ provided by Eq.~(\ref{infdensity}). Our consideration proves that for the infinite density, in contrast with the CLT, the limit of $\alpha\to 2$ of L\'{e}vy walk models with power-law tail $\psi(\tau)\sim \tau^{-1-\alpha}$ is regular up to the infinite constant ($\Gamma(1-\alpha)$ in the calculation above) that is absorbed in the constants. In contrast with the CLT that depends on the statistics of velocity through $\langle v_i v_k\rangle$ only, the infinite density repeats the angular structure of the velocity distribution $F(\bm v)$, see Eq.~(\ref{infdensity}) and thus is less universal \cite{2016,2017}. The infinite density limit is one of the main results of our work and we hypothesize that a similar limit exists for the Lorentz gas.

The infinite density limit (\ref{infdensity}) implies that at large times,
\begin{eqnarray}&&\!\!\!\!\!\!\!\!\!\!\!\!\!\!
P(\bm r, t)\sim \frac{1}{t^{d+1}}I\left(\frac{\bm r}{t}\right),\ \ \bm r\neq 0. \label{infinite}
\end{eqnarray}
The infinite density describes $r\sim t$ that are much larger than the typical $r\sim (t\ln t)^{1/2}$ described by the CLT. Thus when we consider the asymptotic form of $I(\bm v)$ at small $\bm v$,
\begin{eqnarray}&&\!\!\!\!\!\!\!\!\!\!\!\!\!\!
I(\bm v)\!\sim \frac{2{\tilde A}_1}{v^{d+2}}\int_0^{\infty} F(v'{\hat v})v'^{d+1} d v',
\end{eqnarray}
the non-normalizability of $I(\bm v)$ (from which the name of infinite density originates) and hence of $P(\bm r, t)$ in Eq.~(\ref{infinite})) at zero creates no problem because Eq.~(\ref{infinite}) does not hold at small $\bm r$: the normalization integral converges in the region of order $(t\ln t)^{1/2}$ governed by the CLT. This region shrinks to the point $r=0$ after division by $t$ in the considered large times' scaling limit. Thus the limitation $\bm r\neq 0$ in Eq.~(\ref{infinite}) is most significant.

\section{Moments of arbitrary order} \label{dsaio}

So far we have derived the central part of the PDF (where the result is a reduction of \cite{us}) and the tail that extends up to the edge of the PDF's support (given by $Vt$ for Eq.~(\ref{vela}). Thus we do not have the full $P(\bm r, t)$. This is however not an obstacle for finding the moments $\langle r^{\beta}(t)\rangle$ which for $\beta\neq 2$ are described either by the central part or by the tail fully.

We consider the contribution of the infinite density in the moments. We have,
\begin{eqnarray}&&\!\!\!\!\!\!\!\!\!\!\!\!\!\!
\langle |\bm r|^{\beta}\rangle\sim \frac{1}{t^{d+1}}\int I\left(\frac{\bm r}{t}\right)r^{\beta} d\bm r=t^{\beta-1}\int I(\bm v)v^{\beta} d\bm v\nonumber\\&&\!\!\!\!\!\!\!\!\!\!\!\!\!\!
={\tilde A}_1 t^{\beta-1} \int_0^{\infty} dv \int_{v'>v} d \bm v' F(\bm v')\left[2\frac{v'^{2}}{v^{3-\beta}}-\frac{v'}{v^{2-\beta}}\right]\nonumber\\&&\!\!\!\!\!\!\!\!\!\!\!\!\!\!
={\tilde A}_1 t^{\beta-1} \int d \bm v' \int_0^{v'} dv F(\bm v')\left[2\frac{v'^{2}}{v^{3-\beta}}-\frac{v'}{v^{2-\beta}}\right],\label{msd}
\end{eqnarray}
which is a convergent integral provided $\beta>2$. In that case the integrals over the power laws are determined by the upper limits so that the assumption that the moment is determined by the infinite density tail is self-consistent. We find,
\begin{eqnarray}&&\!\!\!\!\!\!\!\!\!\!\!\!\!\!
\langle |\bm r|^{\beta}(t)\rangle\sim \frac{{\tilde A}_1 \beta \langle |\bm v|^{\beta}\rangle}{(\beta-2)(\beta-1)}t^{\beta-1},\ \ \beta>2. \label{largen}
\end{eqnarray}
Setting $\beta=4$ reproduces the fourth moment given by Eq.~(\ref{fourths}). This calculation of the moments is self-consistent since the moments are determined by $r\sim v_t t$ where $v_t$ is the typical velocity that determines the moment $\langle |\bm v|^{\beta}\rangle$.
Similar $t^{\beta-1}$ scaling of the moments with $\beta>2$ was derived for the Lorentz gas in \cite{arm,art,im}. Our calculation predicts that the correction to the infinite density limit are of order $1/t$ and hence Eq.~(\ref{largen}) must be testable numerically.

In contrast for the moments $\langle |\bm r|^{\beta}\rangle$ with $0<\beta<2$ the integral in Eq.~(\ref{msd}) diverges at small $|\bm v|$ or $|\bm r|$. These moments are determined by the central part of the PDF described by the CLT. We have for the velocity model given by Eq.~(\ref{vela}) from Eq.~(\ref{gasq}) that,
\begin{eqnarray}&&\!\!\!\!\!\!\!\!\!\!\!\!\!\!
\langle r^{\beta}(t)\rangle\!\approx\! \int_0^{\infty}\!\! r^{1+\beta}e^{-r^2/\xi^2(t)}\frac{2 dr}{\xi^2(t)}
\!=\!\Gamma\left(1\!+\!\frac{\beta}{2}\right)\xi^{\beta}(t),
\end{eqnarray}
cf. \cite{us}. It is illuminating to use the reduced form of $\xi$ given by (\ref{cl}),
\begin{eqnarray}&&\!\!\!\!\!\!\!\!\!\!\!\!\!\!
\langle r^{\beta}(t)\rangle\!\approx\! \Gamma\left(1\!+\!\frac{\beta}{2}\right)\left(\frac{A_1 }{\langle \tau\rangle}\right)^{\beta/2}\!\left(\ln \frac{t}{\langle \tau\rangle}\!+\!2\zeta\right)^{\beta/2} t^{\beta/2},
\end{eqnarray}
that holds provided $\ln (t/\langle \tau\rangle)\!+\!2\zeta\gg 1$. This formula gives half $\langle r^{2}(t)\rangle$ in $\beta\to 2$ limit and for small $2-\beta$ needs refinements. These can be obtained by direct study of the fractional derivative representation of the moments $\langle r^{\beta}(t)\rangle$ with $0<\beta<2$ in terms of the characteristic function given by the Montroll-Weiss equation. This study is beyond our purposes here.

Finally, $\langle r^{2}(t)\rangle$ cannot be obtained in this way.
This moment however is found simply in the way provided in Sec. \ref{if}. This finishes the study of the moments.

\section{Role of one step: tail and high-order moments} \label{tailk}

In this Section we find the infinite density ${\tilde I}(\bm v)$ for the jump model. We demonstrate that it differs from $I(\bm v)$ found in previous Section and leads to different moments of order higher than two. Difference in only one step of the walk changes these quantities appreciably in contrast with the central part of the PDF and low-order moments considered in Sec. \ref{sum}.

We designate the infinite density of the jump model by ${\tilde I}(\bm v)$. The counterpart of Eq.~(\ref{t1}) for the jump model does not include the factors involving $\bm k$. Correspondingly Eq.~(\ref{s1}) becomes,
\begin{eqnarray}&&\!\!\!\!\!\!\!\!\!\!\!\!\!\!
P\left(\frac{\bm k}{t}, \frac{u}{t}\right)\!\sim 
-{\tilde A}_1\left\langle \left(1\!-\!\frac{i\bm k\cdot\bm v}{u}\right)^2\ln \left(1\!-\!\frac{i\bm k\cdot\bm v}{u}\right) \right\rangle,\nonumber
\end{eqnarray}
where we dropped terms that are either independent of $\bm k$ or proportional to $k^2$. We find,
\begin{eqnarray}&&\!\!\!\!\!\!\!\!\!\!\!\!\!\!
{\tilde I}(\bm v)=-{\tilde A}_1\int\frac{d\bm k}{(2\pi)^d}\frac{du}{2\pi i}\exp\left[i\bm k\cdot\bm w+u\right]
\nonumber\\&&\!\!\!\!\!\!\!\!\!\!\!\!\!\!\times
\left\langle \left(1\!-\!\frac{i\bm k\cdot\bm v}{u}\right)^2\ln \left(1\!-\!\frac{i\bm k\cdot\bm v}{u}\right) \right\rangle.
\end{eqnarray}
We have using,
\begin{eqnarray}&&\!\!\!\!\!\!\!\!\!\!\!\!\!\!
(1\!-\!x)^2\ln(1\!-\!x)\!=\!\frac{3x^2}{2}\!-\!x\!-\!2\sum_{n=3}^{\infty}\frac{x^{n}}{n(n\!-\!1)(n\!-\!2)},
\end{eqnarray}
that,
\begin{eqnarray}&&\!\!\!\!\!\!\!\!\!\!\!\!\!\!
\left\langle \left(1\!-\!\frac{i\bm k\cdot\bm v}{u}\right)^2\ln \left(1\!-\!\frac{i\bm k\cdot\bm v}{u}\right) \right\rangle.
\nonumber\\&&\!\!\!\!\!\!\!\!\!\!\!\!\!\!
=-\sum_{n=2}^{\infty} \left\langle  \frac{(-1)^n(\bm k\cdot\bm v)^{2n}}{u^{2n}(2n-2)(2n-1)n} \right\rangle-\frac{3k^2\langle v_i^2\rangle}{2u^2}.
\end{eqnarray}
Thus,
\begin{eqnarray}&&\!\!\!\!\!\!\!\!\!\!\!\!\!\!
{\tilde I}(\bm w)={\tilde A}_1\int\frac{d\bm k}{(2\pi)^d}\exp\left[i\bm k\cdot\bm w\right]\left\langle {\tilde G}(\bm k\cdot\bm v) \right\rangle
\nonumber\\&&\!\!\!\!\!\!\!\!\!\!\!\!\!\!
{\tilde G}(y)=\sum_{n=2}^{\infty}  \frac{(-1)^n y^{2n}}{(2n-1)!n(2n-2)(2n-1)}.\label{tr}
\end{eqnarray}
We observe that $G=y{\tilde G}'/2$ so that (we use summation convention over repeated indices),
\begin{eqnarray}&&\!\!\!\!\!
I(\bm w)=\frac{{\tilde A}_1}{2} \int\frac{d\bm k}{(2\pi)^d}\exp\left[i\bm k\cdot\bm w\right]\left\langle k_i\frac{\partial}{\partial k_i} {\tilde G}(\bm k\cdot\bm v) \right\rangle\nonumber\\&&\!\!\!\!\!
=-\frac{{\tilde A}_1}{2} \int\frac{d\bm k}{(2\pi)^d}\left\langle  {\tilde G}(\bm k\cdot\bm v) \right\rangle \frac{\partial}{\partial k_i}\left(-i\frac{\partial}{\partial w_i}\exp\left[i\bm k\cdot\bm w\right]\right)
\nonumber\\&&\!\!\!\!\!=-\frac{1}{2}\frac{\partial \left(w_i{\tilde I}(\bm w)\right)}{\partial w_i}=-\frac{1}{2w^{d-1}}\frac{\partial (w^d{\tilde I})}{\partial w}.
\end{eqnarray}
Switching to variable $\bm v$ instead of $\bm w$,
\begin{eqnarray}&&\!\!\!\!\!
{\tilde I}(\bm v)=\frac{2}{v^d}\int_v^{\infty} v'^{d-1} I(v'{\hat v})dv'=\frac{2{\tilde A}_1}{v^d}\int_v^{\infty}dv' \int_{v''>v'}d v''
\nonumber\\&&\!\!\!\!\! F(v''{\hat v})v''^{d-1} \left[2\frac{v''^{2}}{v'^3}-\frac{v''}{v'^{2}}\right]=\frac{{\tilde A}_1}{v^{d-1}}\int_v^{\infty}F(v'{\hat v})v'^{d-1}
\nonumber\\&&\!\!\!\!\! \left(2\frac{v'^{2}}{v^3}-2\frac{v'}{v^2}\right)dv'.
\end{eqnarray}
where we used Eq.~(\ref{infdensity}). Similar formula in the case of one-dimensional L\'{e}vy walk with tail $\psi(\tau)\sim \tau^{-1-\alpha}$ was derived in \cite{infden}.

We see by comparison with Eq.~(\ref{infdensity}) that the infinite density tails of the jump and velocity (our LW) models differ in the last term in the integrand. This term is irrelevant for dispersion $\langle r^2(t)\rangle$ in the leading order in accord with the previous observation that the dispersions of the two models agree in the leading order at large times.
The corrections however already differ, see Eqs.~(\ref{growth01}) and (\ref{se}). For moments of order $\beta>2$ we find as in Eq.~(\ref{msd}),
\begin{eqnarray}&&\!\!\!\!\!\!\!\!\!\!\!\!\!\!
\langle |\bm r|^{\beta}\rangle\sim
2{\tilde A}_1 t^{\beta-1} \int\!\! d \bm v' \int_0^{v'}\!\! dv F(\bm v')\left[\frac{v'^{2}}{v^{3-\beta}}\!-\!\frac{v'}{v^{2-\beta}}\right].
\end{eqnarray}
We find that the counterpart of Eq.~(\ref{largen1}) for the jump model is,
\begin{eqnarray}&&\!\!\!\!\!\!\!\!\!\!\!\!\!\!
\langle |\bm {\tilde r}|^{\beta}(t)\rangle\sim \frac{2{\tilde A}_1  \langle |\bm v|^{\beta}\rangle}{(\beta-2)(\beta-1)}t^{\beta-1},\ \ \beta>2. \label{largen1}
\end{eqnarray}
Thus the moments of velocity model are larger than the moments of the jump model by a constant factor of $\beta/2>1$. Identical result for the moments' ratio was found in \cite{infden} for one-dimensional L\'{e}vy walk with tail $\psi(\tau)\sim \tau^{-1-\alpha}$ where $\alpha<2$. This confirms again the regularity of $\alpha\to 2$ limit for the tail of the distribution (up to the constants' redefinition considered previously).

\section{Conclusions}

We demonstrated that predictions of the L\'{e}vy walk model of the Lorentz gas agree with all the properties of the gas known so far. Thus it can be hoped that further transfer of the LW's properties to the LG is possible. The highest interest is in numerical tests of the refined CLT (which was done for $R=0.4$ and is thus less urgent) and the tail of the distribution that so far was not measured. The form provided by the LW seems to be almost necessarily true for the LG however the numerical proof is necessary.

On the theoretical size, the immediate goal is the rigorous refinement of the Bleher CLT for inclusion of the normal diffusion component of the Gaussianity and the next-order logarithmic term. It seems that the steps of \cite{bleher} used with the refined CLT of Sec. \ref{fast} for discrete sums go through. This leads to the refined CLT however a detailed study is necessary.

We demonstrated that the growth of the dispersion and the fourth moment of the particle's displacement are determined by the statistics backward recurrence time $B_t$. It is of interest to see if $B_t$ can also be of use for the LG.

Our work seem to complete the theoretical study of the properties of the LW model of the LG started in \cite{us}. We hope that tests will confirm the theory fully, thus providing us with a good understanding of this classical physical model.

%
%
%

\section{Acknowledgements}

I. F. cordially thanks E. Barkai for the introduction to the area of L\'{e}vy walks and is grateful to S. Denisov and L. Zarfaty for helpful discussions. P. D. thanks the Villum foundation, project no. $17470$, for supporting I.F.'s research visit in Copenhagen.

{}

\end{document}